\RequirePackage{fix-cm}
\documentclass[twocolumn,epjc3]{svjour3}           
\smartqed  
\usepackage{graphicx}
\usepackage{appendix}
\usepackage{siunitx}
\begin{document}

\title{Reconstructing the arrival direction of neutrinos in deep in-ice radio detectors}

\titlerunning{Radio arrival direction}        

\author{Ilse Plaisier\thanksref{e1,fau, DESY}         
        \and
        Sjoerd Bouma\thanksref{e1,fau} 
        \and 
        Anna Nelles\thanksref{e1,DESY, fau}
}


\institute{Erlangen Center for Astroparticle Physics (ECAP), Friedrich-Alexander-Universit{\"a}t Erlangen-N{\"u}rnberg, Nikolaus-Fiebiger-Straße 2, 91058 Erlangen, Germany \label{fau}
            \and
            Deutsches Elektronen-Synchrotron DESY, Platanenallee 6, 15738 Zeuthen, Germany \label{DESY}                    
}

\thankstext{e1}{e-mail: ilse.plaisier@desy.de, sjoerd.bouma@fau.de, anna.nelles@desy.de}
\date{Received: date / Accepted: date}

\maketitle

\begin{abstract}
In-ice radio detectors are a promising tool for the discovery of EeV neutrinos. For astrophysics, the implications of such a discovery will rely on the reconstruction of the neutrino arrival direction. This paper describes a first complete neutrino arrival direction reconstruction for detectors employing deep antennas such as RNO-G or planning to employ them like IceCube-Gen2.
We will didactically introduce the challenges of neutrino direction reconstruction using radio emission in ice, elaborate on the detail of the algorithm used, and describe the obtainable performance based on a simulation study and discuss its implication for astrophysics. 

\end{abstract}

\section{Introduction}
\label{sec:intro}
Neutrinos with energies up to \SI{10}{PeV} have been detected numerously and IceCube has discovered a component of astrophysical neutrinos above the atmospheric neutrino background \cite{IceCube:2014stg}. These extraterrestrial neutrinos are expected to be created in extreme cosmic sources that accelerate charged particles, cosmic rays, to high energies, which produce secondary particles in their interactions with ambient matter and photon fields: neutrinos e.g.~\cite{Margolis:1977wt,Berezinsky:1969erk,ParticleDataGroup:2022pth}. Several point sources are revealing themselves as source candidates for neutrinos, and therefore of the parent cosmic rays, either by an excess of neutrino events or through a coincident detection with multi-messenger particles \cite{IceCube:2018cha,Stein:2020xhk,NGC}. However, the first neutrino source still needs to be identified at $5 \sigma$ significance. As neutrinos typically carry $\mathcal{O}(1/20)$ of the energy of the parent nucleon \cite{Mucke:1999yb}, a different yet undiscovered population of neutrinos will be the result of the interactions of ultra-high energy (UHE) ($>$ EeV) cosmic rays, for which the energy spectrum is accurately established \cite{PierreAuger:2010gfm,Verzi:2017hro}. Radio detectors are proposed for the detection of the resulting UHE neutrinos. Among these UHE neutrinos also \textit{cosmogenic neutrinos} are expected, which result from the interaction of cosmic rays with photon fields during their propagation towards Earth \cite{Berezinsky:1969erk}.

The currently largest neutrino detectors use optical Che\-ren\-kov light stemming from the secondary particles that are created when the neutrino interacts (e.g.\ \cite{IceCube:2014stg,KM3Net:2016zxf}). Since this light has a propagation length of $\mathcal{O}(100)$ m due to scattering and absorption in the relevant dense media \cite{OpticalProperties2006}, effective volumes are limited, as the volume is required to be densely instrumented. The secondary particle showers also emit nanosecond-duration radio pulses, which can travel for $\mathcal{O}$(\SI{1}{km}) in dielectric media such as ice. Therefore sparse radio arrays can be built to cover large volumes. Where in-ice radio detectors are in principle sensitive to the highest energies ($>$10 PeV), they rapidly lose sensitivity below 100 PeV, due to the irreducible thermal noise of the antennas and the surroundings, which prevents the detection of very low amplitude signals.

This work focuses on radio detectors using ice as the detection medium. The technique is used for in-ice radio neutrino detectors in the pilot arrays ARIANNA \cite{Anker:2019rzo} and ARA \cite{Allison:2011wk}, in RNO-G, currently being constructed in Greenland \cite{RNO-G:2020rmc}, and the proposed radio component for the next generation of IceCube, IceCube-Gen2 \cite{IceCube-Gen2:2020qha}. RNO-G is scheduled to be completed with 35 stations in 2026 and will be complementary to IceCube-Gen2 in the northern hemisphere. It will be the first in-ice radio array of sufficient scale to have a realistic chance of discovering the first \textit{radio neutrino}. Such a discovery will be facilitated by a good angular resolution. Its impact for astrophysics will be also enhanced by a reconstruction of the neutrino arrival direction \cite{Fiorillo:2022ijt}.  Besides aiding in the identification of neutrino sources, a good resolution in zenith helps to firmly establish the diffuse UHE neutrino flux, as not only the flux but also the neutrino cross section is not yet known at these energies \cite{Valera:2022wmu}.
IceCube-Gen2, with a planned effective volume of more than 1600 km$^3$ sr, will reach the discovery space of almost all predictions of UHE neutrino fluxes, e.g.~\cite{Fang:2013vla,Padovani:2015mba,Heinze:2019jou,Rodrigues:2020pli,Muzio:2021zud}.   

In-ice radio neutrino detectors typically consist of a cluster of in-ice antennas forming a self-triggering \textit{station}, with distances between stations of $\mathcal{O}(1)$ km such that there is little overlap in effective volume for individual stations. It is estimated that only 10\% of the triggering neutrinos will be detectable in two stations. Radio neutrino stations are typically classified as \textit{shallow stations}, \textit{deep stations}, and \textit{hybrid stations}. Shallow stations contain broadband antennas buried \SIrange{1}{3}{m} below the ice surface. They are limited in effective volume, because the gradually changing refractive index with depth (due to the increasing ice density) results in parts of the ice from which the radio signal is not able to reach the detector. Deep stations contain deep in-ice holes (down to depths of \SIrange{100}{200}{m}) in which \textit{strings} of antennas are installed. They reach down to the bulk ice, providing a large effective volume, but these holes are narrow, $\mathcal{O}$(\SI{30}{cm}), and therefore limit the geometry of the antennas. Hybrid stations are a combination of a deep and shallow station. Both RNO-G and IceCube-Gen2 will have hybrid stations.

\subsection{Previous work}
A number of reconstruction algorithms for different aspects of radio particle detectors have been developed and resolutions are highly dependent on the specific station design and detection medium. 

The reconstruction of the signal arrival direction is best studied. It depends primarily on how well the antenna position and the instrument timing is known, as well as the assumed waveform model. In air showers, direction resolutions of better than 0.5$^\circ$ are common, if many antennas detect the signal and a hyperbolic waveform fit can be used, as shown for e.g.\ LOPES \cite{LOPES:2021ipp} and LOFAR \cite{Corstanje:2014waa}. For in-ice experiments, the ARIANNA experiment (shallow antenna installation) reported an angular resolution of $<1^\circ$ for a pulser lowered into the ice \cite{ARIANNA:2020zrg}. The ARA collaboration (deep antennas) has reported a similar angular resolution for pulsers installed at large depths \cite{Allison:2011wk}.

Some aspects of reconstructing the signal polarization are also relatively well studied. Air shower experiments have shown that the statistical uncertainty on the angle of polarization is inversely proportional to the signal-to-noise ratio approaching $0.1^{\circ}$ for clear signals \cite{Schellart:2014oaa}. Additional systematic uncertainties may stem from the antenna modeling, but were estimated to reach better than 10\% \cite{PierreAuger:2014ldh}. The reconstructions even allowed for the measurement of the small circular component in the in-air emission, stemming from the time difference between geomagnetic and Askaryan emission \cite{Scholten:2016gmj}.

The situation in ice is more complex for two reasons:  
Glacial ice shows a gradient in the index of refraction that significantly curves trajectories and the ice is known to exhibit birefringent properties that will influence how well the polarization can be reconstructed \cite{Heyer:2022ttn,Connolly:2021cum}. So far, reconstructions have taken into account the bending, but mostly ignored the birefringence. Second, in the reconstruction of the electric field, antennas at different positions have to be combined rather than using dual-polarized antennas like in air, since antenna sizes are restricted due to installation constraints. This adds additional timing and positioning uncertainty. 

For shallow antennas a polarization resolution of order 1$^\circ$ was obtained  based on cosmic-ray data, propagating only briefly through the ice, but in excellent agreement with simulations \cite{Arianna:2021lnr}. ARIANNA also reported a resolution of 3$^\circ$ for in-ice pulser data, including emitter uncertainties, albeit only probing a small range of polarization angles with high signal-to-noise data \cite{ARIANNA:2020zrg}. 

The ARA experiment has reported on a number of indications for birefringence \cite{Allison:2017jpy,Allison:2019rgg,Besson:2021wmj}. Their data probe much longer distances and more polarization angles, but a complete model of the signal propagation of neutrino pulses is still outstanding, preventing the full electric-field reconstruction for the polarization reconstruction of pulser signals.  

Aside from experimental evidence, several simulation studies have been performed to study reconstruction capabilities of in-ice radio stations. Machine learning algorithms show a lot of promise for neutrino angular reconstruction. For shallow antennas, a space angle difference of roughly $4^{\circ}$ has been found \cite{Glaser:2022lky}. Similar results have been reported for ARA, albeit less studied in detail at this point \cite{ARA:2021bss}.  

The neutrino energy resolution for deep antennas has been studied in \cite{Aguilar:2021uzt}. The presented method uses Information Field Theory (IFT) \cite{Welling:2021cgl} to reconstruct the electric field at the antennas. The amplitude of the measured electric field pulses scale linearly with the energy and inversely with the distance to the interaction vertex. Therefore, the energy reconstruction is highly dependent on the distance to the interaction point of the neutrino, i.e.\ the \textit{vertex position}.

Due to the in general low amplitude of the signal pulses,  standard methods that unfold the voltage data with the detector response to obtain the electric field typically overestimate certain signal parameters and perform worse than methods where noiseless waveforms are matched with the data. Examples are the IFT approach and the \emph{forward-folding approach}, where noiseless voltage waveform expectations are obtained by forward-folding an electric field assumption through the ice and detector response \cite{Glaser_2019}. Forward-folding will also be used as the basis of this work. It has previously been applied to shallow antennas, resulting in a angular resolution of 3$^\circ$ \cite{Gaswint:2021smu}. 

Besides the applicability to low amplitude events, the model-dependency of our method has the advantage (as compared to the current implementation of IFT) that it takes into account the different electric fields at different antenna positions, i.e.\ it \textit{knows} where the antennas are. 

After a didactic overview of neutrino angular reconstruction in section \ref{sec:didactics}, explaining the complexity of pinpointing the neutrino arrival direction due to the shower geometry, we will describe the algorithm and reconstruction approach in section \ref{sec:Reconstruction}. We show the performance of the algorithm and the resulting angular resolution in section \ref{sec:performance}. We will also discuss the impact of different selection cuts on the obtained resolutions. Furthermore, we will show event contours of typical events, as well as the point spread function, and discuss the implications for point-source studies. Our conclusions will be summarized in section \ref{sec:conclusions}.

\section{Radio detection of neutrinos}
\label{sec:didactics}

In-ice radio detectors measure the radio emission stemming from particle showers induced by neutrinos interacting in vast ice volumes. The radio emission comes from an accumulated negative charge excess which varies over time due to the shower development, the \textit{Askaryan effect} \cite{Askaryan}. It results in coherent nanosecond long radio pulses, emitted on a cone with its maximum amplitude at $\approx$ 56$^\circ$ (for deep ice in Greenland), the Cherenkov angle $\Theta_{C}$.

Due to the cone, the signal emission direction is not equal to the neutrino arrival direction. In addition, due to the varying index of refraction, the  direction  of  the  signal  changes  during propagation such that the emission direction is not the same as the direction when the signal arrives 
at the detector. However, 
a measure of the \textit{signal direction}, the angle between the radio emission and the neutrino direction (the \textit{viewing angle}), and the \textit{polarization} of the electric field are sufficient to fully constrain the neutrino direction. This leads to a complex number of steps involved to reconstruct the neutrino direction, which will be elaborated in the following sections. 

\subsection{Connection of neutrino direction and signal parameters}
\label{subsection:parameters}
As the particle shower propagates at a speed close to $c$, and hence faster than the speed $v = \frac{c}{n}$ of the electromagnetic waves, with $n$ denoting the refractive index, radio waves arrive simultaneously at an observer at $\Theta_{C}$ from the shower axis. This observer sees fully coherent radio waves. For slightly off-cone angles, the coherence for high frequencies is lost, resulting in a lower cut-off frequency and rapidly decreasing signal amplitude. Therefore, the electromagnetic waves can only be observed on a cone around the shower axis with opening angles close to the Cherenkov angle $\Theta_{C}\pm 10^\circ$ for detectors sensitive to MHz frequencies.

\begin{figure*}
    \centering
    \includegraphics[width=0.9\textwidth]{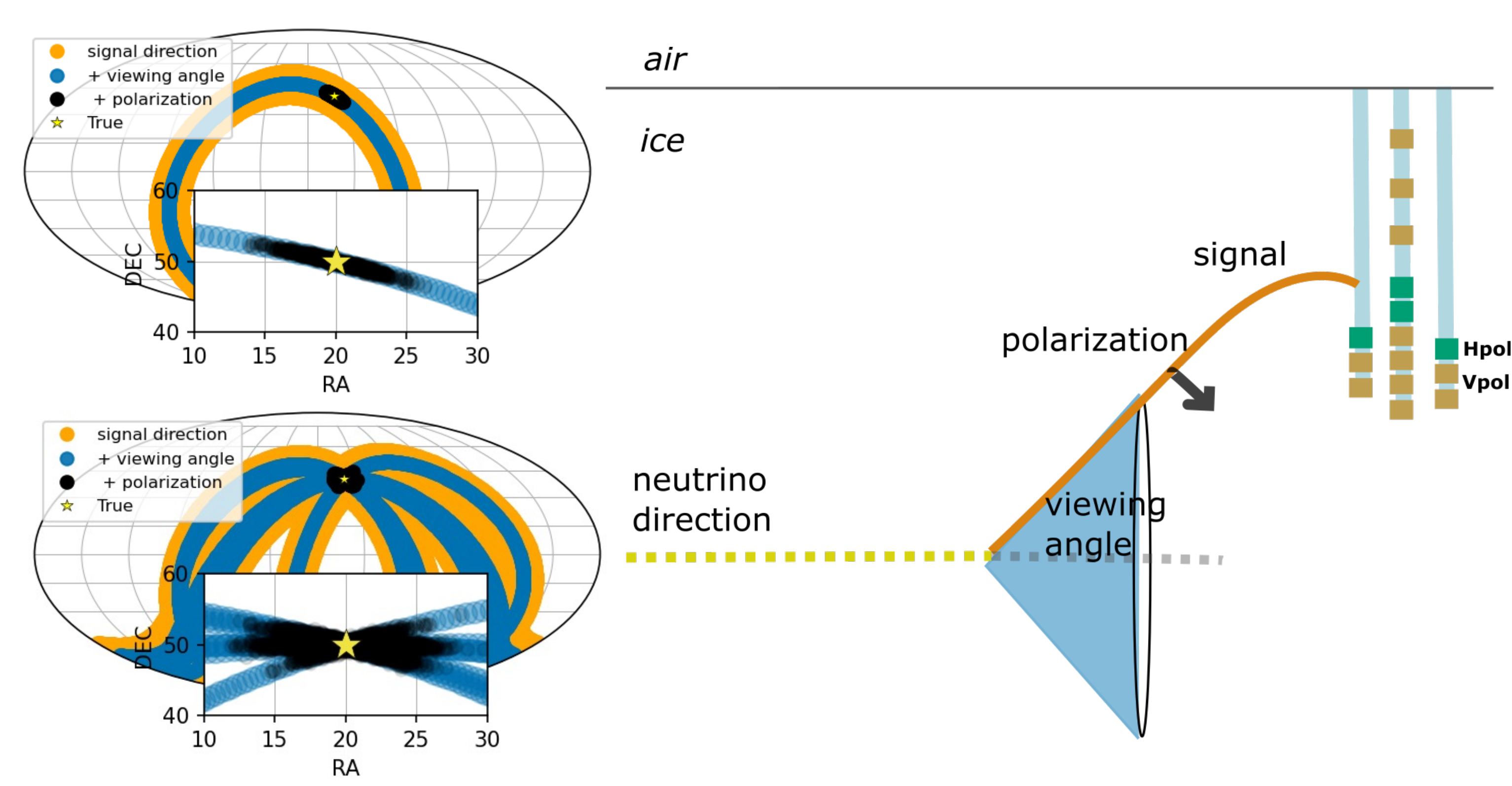}

    \caption{Illustration of the shower geometry and the radio cone on which the signal is typically detectable (right, not to scale), as well as the impact on the reconstruction of a single event (top left) and the point spread function (bottom left). The point spread function is shown based on events with a number of different viewing angle and polarization combinations to illustrate the difference with the single event reconstruction. The radio signal is observed at the detector under the viewing angle, i.e.\ the angle between the shower axis and the radio signal, which is typically a few degrees off from the Cherenkov angle. The electric field is polarized towards the shower axis due to the Askaryan effect. A measurement of the signal direction, viewing angle and polarization angle results in the reconstruction of a single neutrino as illustrated in the top left figure. Multiple neutrinos from a point source are reconstructed as shown in the bottom left figure.}
    \label{fig:educative_figure}
\end{figure*}

It also means that the frequency content of the observed pulses contains information on the observed viewing angle of the radio emission with respect to the shower axis. 
The negative charge excess in the shower front, leads to a polarized electric field with its electric field vector pointing towards the shower axis. Therefore, combined measurements of the signal emission direction, viewing angle, and polarization suffice to reconstruct the direction of the original neutrinos. This is illustrated in Figure \ref{fig:educative_figure} left. In reality, the  method reconstructs the direction of the particle 
showers. However, the shower is very strongly forward-boosted already at 1 TeV \cite{Chiarusi:2010qif}, such that at the energies relevant here the angular deviation between the shower and the original neutrino is completely negligible. 

\subsection{Modeling of neutrino radio signals}
\label{subsection:signal_model}
Radio emission from showers can be calculated from first principles in microscopic particle simulations, e.g.\ \cite{Huege:2013vt,Alvarez-Muniz:2011ref}. This is, however, computationally expensive and only needed, if signal parameter accuracies below 10\% are needed. For neutrino detection, where large uncertainties stem from the propagation in ice, parameterizations are currently sufficient. Modeling the electric field expected from the Askaryan emission from a particle shower is best done using a semi-analytical model, that calculates the time domain waveform directly from the charge-excess distribution in the shower via convolution with a form factor that only depends on the shower type, i.e.\ hadronic and electromagnetic showers \cite{Alvarez-Muniz:2011wcg}. The model shows good agreement ($3\%$) with Monte Carlo codes. At the cost of a loss of accuracy, simpler (and thus faster) parameterizations are also available. 

In addition to the parameters discussed in section \ref{subsection:parameters}, the Askaryan emission also depends on the type of particle shower initiated by the neutrino. For $\nu_\mu$, $\nu_\tau$, and neutral current interactions from $\nu_e$, only a hadronic shower is created, but for $\nu_e$ charged current interactions, the electron additionally gives rise to an electromagnetic shower. At ultra-high energies, electrons are subject to the LPM effect, resulting in a more irregularly shaped shower profile than hadronic showers \cite{Landau:1953um,Migdal:1956tc}, and the presence of multiple showers can cause interference at the detector. Accordingly, the reconstruction of these events is expected to be more challenging. The algorithm outlined in this article was therefore designed for hadronic events. We do, however, include all events in our simulation, and discuss the impact on performance in section \ref{sec:performance}.

The current state-of-the-art neutrino simulation code, NuRadioMC \cite{Glaser:2019cws}, provides an interface to most of the models. NuRadioMC modularizes the simulation problem into event generation (neutrino vertices), signal generation (chosen from a parameterization), signal propagation, and signal detection (flexibly including various detector types). NuRadioMC is open source and under continuous development, adding for example more complex propagation models and better descriptions of the detectors. All simulations described here are performed with NuRadioMC.

\subsection{Signal propagation}
\label{subsection:signalpropagation}
NuRadioMC uses the decoupling of signal propagation and signal generation, i.e.\ for a given neutrino interaction position the path traveled through the detector is calculated using ray tracing. The electric field as propagated on this path and observed at the antenna is then calculated. The ray tracing approximation holds when the wavelength is much larger than the relevant features in the ice. In most cases this is sufficient, i.e.\ the refractive index is slowly increasing and the ice is homogeneous. 

The refractive index of polar ice is increasing with depth, reaching a constant value of 1.78 for deep ice at $\approx$ \SI{100}{m} (Greenland). The resulting changing speed of the radio waves causes the waves to bend towards regions of higher refractive index, so that signals from neutrino vertices below an observer can also arrive refracted from above. At an air-ice, ice-water or ice-rock interface, radio waves are (partly) reflected. During propagation, the radio waves are attenuated, the attenuation length of the medium is both depth and frequency dependent. All of these features are taken into account in NuRadioMC, resulting in so-called \emph{direct}, \emph{refracted}, and \emph{reflected} signal propagation paths.  
Horizontal propagation of signals near the ice-air surface or due to density fluctuations in the firn \cite{PhysRevD.98.043010,Barwick:2018rsp} are currently not taken into account.

Recent data has shown that polar ice can be \textit{birefringent}, i.e.\ the ice crystal orientation causes the refractive index to be dependent on the direction and polarization of the radio waves which results in a time-delay for different polarization components. Work is ongoing in understanding and simulating this effect e.g.~\cite{Heyer:2022ttn,Jordan:2019bqu,Connolly:2021cum}, but the modeling is currently not mature enough to include this in our simulation study.
Furthermore, layers in the ice with density fluctuations can cause partial reflection of the radio waves. These layers are observed at South Pole \cite{Besson:2021wmj} and Greenland \cite{Aguilar:2022ijn}. Also their effect on the angular reconstructing has to be studied in future work.

\subsection{Detector response}
\label{subsection:detector_response}
Deep in-ice stations typically contain two kinds of antennas to measure orthogonal electric field components, designed to be predominantly sensitive to either the vertical polarization (Vpol) or the horizontal polarization (Hpol). The Vpol is designed to be high gain and broadband in the $\mathcal{O}(100-600)$ MHz range, such that information on the viewing angle can be extracted from the frequency content. The Hpol design, typically lower in gain due to the geometry constraints by the borehole, is designed to have sufficient overlap in the frequency content with the Vpol to aid polarization reconstruction.

In-ice neutrino radio arrays are designed to optimize effective volumes, hence the independent triggering stations are located on a sparse grid. The bulk of neutrinos will trigger only a single station, therefore each station is designed to be capable of reconstructing neutrino properties. A schematic of the deep station used in this work is shown in Figure \ref{fig:educative_figure} (right). It contains three \textit{strings} of antennas at lateral distances of \SI{35}{m}. The holes are drilled to such a depth that the deepest point of the strings is below the firn, to maximize the effective volume. At the bottom of the \textit{power string} closely spaced Vpols (\SI{1}{m}) are located. By beamforming the waveforms in these antennas,  which is an effective noise reduction, they function as a low-threshold trigger called the \textit{phased array}. Furthermore, the power string is equipped with Hpols located right above the phased array, and Vpols at larger distances. The other two \textit{helper strings} have Vpols and an Hpol located at the lower end of the string. 

Simulations take into account both the antenna locations, as well as their complex performance (gain and group delay) to incoming signals. Also, the effect of the electronics chain (amplifiers, cable losses, digitization, triggering) is simulated, to be able to study the reconstruction efficiency. The station modeled here, closely resembles an RNO-G station as currently under construction.

\subsection{Event contour and point spread function}
\label{sec:contours}
Due to the number of steps involved in reconstructing the neutrino arrival direction for radio neutrino detectors, there is a distinct difference between the shape of a single event contour and the point spread function (PSF). As will be elaborated further in detail, both should only be approximated in two dimensions. This is illustrated in Figure \ref{fig:educative_figure}. 

A measure of the signal arrival direction restricts the neutrino direction to a broad band, because the amplitude at the detector will only be large enough to be detected for angles a few degrees off the Cherenkov cone. A viewing angle measurement narrows this band. An additional polarization measurement restricts the location of this band. This results in ellipse like single-event uncertainties for good quality events.

Reconstructing the polarization is most challenging for deep in-ice radio detectors.  The design of the antennas is constrained by the borehole diameter, typically $\approx$ 30 cm wide, due to the available drills. This makes designing antennas sensitive to the horizontal signal component (Hpol antennas) more challenging than the ones sensitive to the vertical component (Vpol antennas). This then results in a significantly smaller gain for the Hpols, with them often measuring no detectable signal. Therefore, the main contribution to the angular resolution is the uncertainty in the polarization measurement resulting in ellipse-like uncertainty contours with very large axis ratios.

A neutrino source observed with an in-ice radio station will be seen smeared out due to the reconstruction uncertainties, which results in the sum of ellipse-shape contours. Due to the geometry of the shower and the corresponding radio-cone, and the fact that the Vpol antennas are used for triggering, the range of interaction positions of the neutrino in the ice (thus signal arrival directions) that can trigger a station is limited. Since the orientation of the ellipse is determined by the signal arrival direction, this results in orientations of the ellipse that are forbidden, resulting in a 'bow-tie'-shaped PSF.

\section{Algorithm for direction reconstruction}
\label{sec:Reconstruction}

It has been first shown in \cite{Glaser_2019} that forward-folding is a suitable reconstruction tool for radio detectors. In the forward-folding approach, an analytic description of the electric field is forward-folded through the propagation and system response to obtain waveforms of the voltage received at every antenna. These waveforms are compared with the recorded voltage data and a test statistic is minimized to obtain the best fitting signal parameters. This approach has shown to work better for reconstructing low signal amplitude radio pulses than unfolding the detector response. Due to the contribution of noise in the voltage data, unfolding the (noiseless) detector response systematically overestimates the signal contribution when the detector response expects a low value.

For reconstructing the neutrino direction, the forward-folding approach of \cite{Glaser_2019} has been extended to use antennas with large distances to each other, where it no longer holds that the same signal is measured. 
This extension has the benefit that using more antennas in the same reconstruction effectively reduces the noise contribution and therefore improves the obtainable resolution. Furthermore, combining antennas across the entire station results in a more accurate measurement of the polarization due to the vertical and horizontal spatial separation than obtainable by only using nearby antennas.

\subsection{Electric field model} 
\label{efield}

For reconstructing the radio pulses an analytical description of the electric field as provided by \cite{Alvarez-Muniz:2010hbb} is used which gives a parameterization of the electric field in the frequency domain, where the amplitude is linearly rising with frequency up to a cut-off frequency ($\approx$ 1.2 GHz for on cone angles), which reduces for off-cone angles.
The parameterization depends on the viewing angle assuming the emission is coming from the shower maximum, i.e.\ the point which most of the emission originates from.  
Besides the viewing angle, the electric field parameterization depends only on the shower energy $E_{\textrm{sh}}$ which determines the signal amplitude.
Since we use a parameterization in the frequency domain, signals coming from different parts of the shower arrive at different times, i.e.\ the phase, is not taken into account. Consequences for the reconstruction procedure are explained in section \ref{sec:approach}. It should be noted that the parameterization used to reconstruct the data, is simpler than the one used to model that data (see section \ref{subsection:signal_model}), which should make the approach more robust towards slight mismodeling of the signal.

Since the electric field model depends on the viewing angle, the interaction point of the neutrino is required to compute the viewing angle under which the radio emission is seen at the antenna for a given neutrino direction. Therefore, reconstructing the interaction point is the first step in the reconstruction procedure. 

\subsection{Neutrino vertex reconstruction}
We use the vertex reconstruction algorithm, which effectively reconstructs the shower maximum, as described in \cite{Aguilar:2021uzt}. As there have been no major changes to the algorithm, we only provide a brief summary below, and refer to \cite{Aguilar:2021uzt} for a more detailed description.

The relative pulse arrival times in the different antennas depend on the vertex position. Therefore, the vertex position can be triangulated by shifting the voltage traces depending on the hypothesized vertex position, and maximizing the resulting antenna-to-antenna pairwise correlation. In fact, rather than correlating the voltage traces directly with each other, they are first correlated with a neutrino template, reducing the effect of accidental correlations in the noise in each antenna.
In addition, a bandpass filter with a passband of \SIrange[]{95}{300}{MHz} is used before performing the correlation. The pulse shape in this band is determined mostly by the antenna and amplifier response, and varies only slightly depending on the arrival direction and viewing angle. 

One additional challenge with respect to traditional triangulation arises due to the presence of multiple ray paths between vertex and antenna position, as explained in section \ref{subsection:signalpropagation}. This is taken into account by testing the time shifts for each possible ray type ('direct' or 'refracted'/'reflected') for each antenna. Finally, one additional subtlety arises due to the fact that the emission does not originate from the neutrino interaction vertex, but from the shower, whose core extends over the order of tens of meters (or more, for electromagnetic showers affected by the LPM effect \cite{Landau:1953um,Migdal:1956tc}). Therefore effectively the position of the shower maximum is reconstructed and not the vertex, which has been accounted for in the following steps.

\subsection{Reconstruction algorithm}
\label{sec:approach}

\begin{figure*}
	\centering
	\includegraphics[height=.25\textheight]{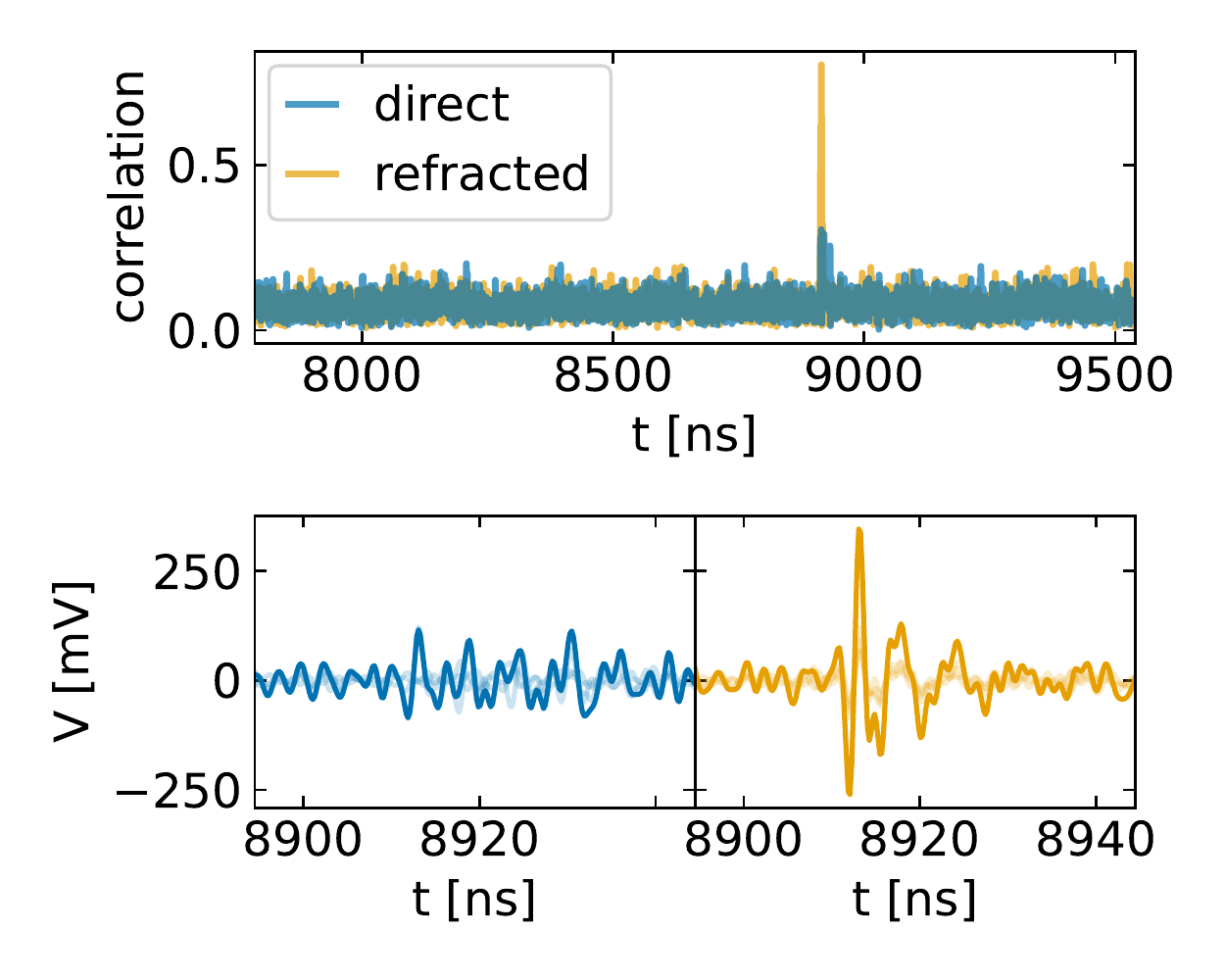}
	\includegraphics[height=.25\textheight]{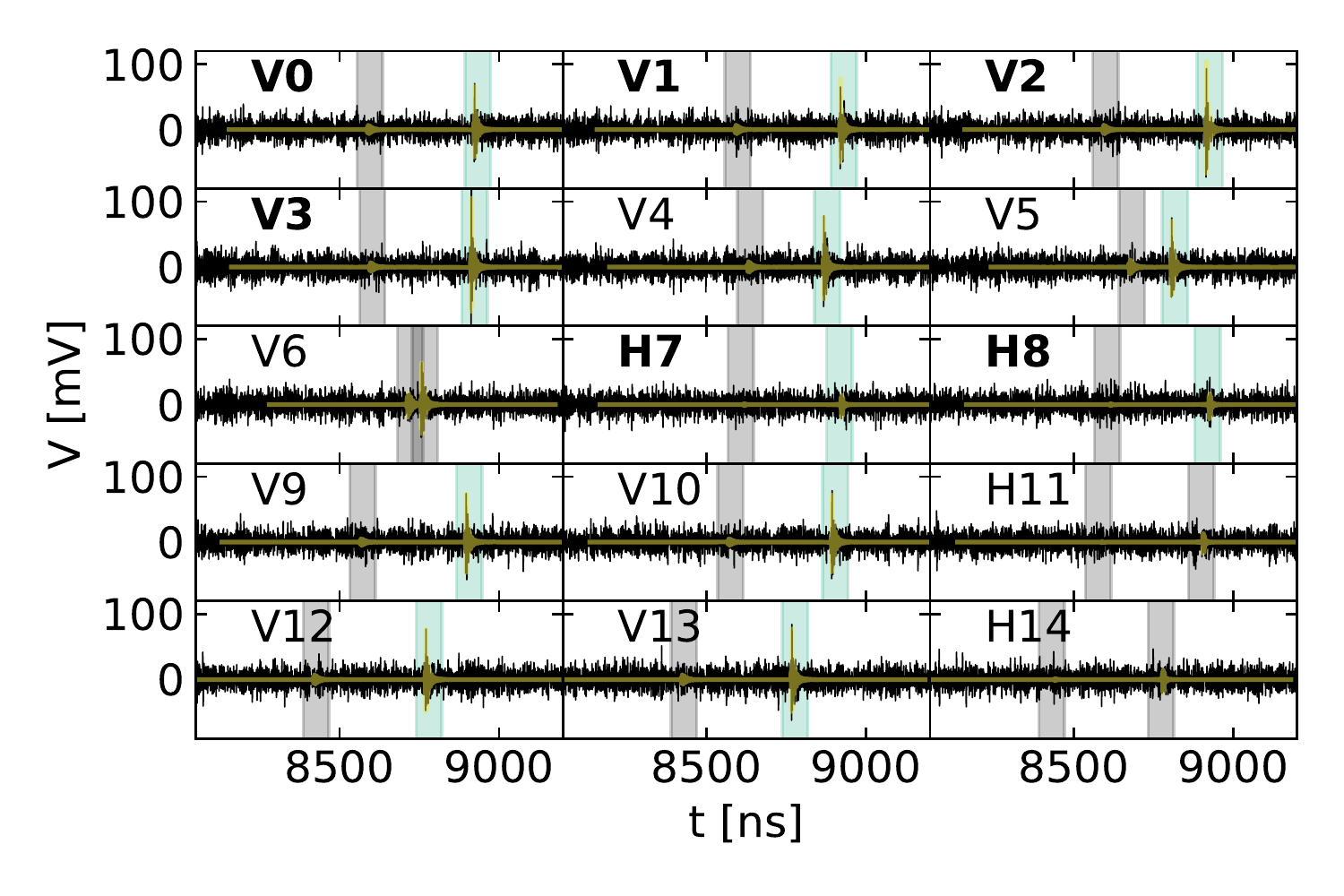}
	\caption{The pulse selection procedure. Left, top: the total template correlation for the phased array antennas for each ray-type hypothesis ('direct' or 'refracted'/'reflected'), indicating here a better correlation for refracted. Left, bottom: The individual as well as coherently-summed (beamformed) pulses for each ray type. Right: Window selection for both ray types. The pulse windows to include in the fit are selected based on their expected locations relative to the pulse arrival times in the phased array, and a cut on the peak-to-peak SNR. Included pulse windows are highlighted in green, excluded ones in gray. The pulses corresponding to the 'phased array cluster' (shown in boldface), consisting of the phased array and adjacent Hpol antennas, are always included in the fit. }
	\label{fig:pulse_selection}
\end{figure*}

Although the reconstructed vertex position provides the expected relative time delays between the different antennas, the absolute pulse arrival time is needed in addition in order to determine the pulse windows to include in the fit. Pulse windows are used to limit the options for accidental correlations. The absolute time is determined by the ray-type selection algorithm. As in the case of the vertex reconstruction algorithm, a template is correlated with the voltage traces for each ray-type hypothesis, and the dominant ray type and its arrival time are determined by looking at the maximum total resulting correlation. This is illustrated in Figure \ref{fig:pulse_selection}.

Unlike in the vertex reconstruction algorithm, here only the four deep Vpol antennas in the phased array are used. As they form the trigger of the experiment, the likelihood of observing a clear (when beamformed) pulse here is highest. In addition, their proximity to each other reduces the relative timing error due to differences in viewing angle or uncertainties in e.g.\ the ice model.

Once the pulse arrival time and the ray type in the phased array have been identified, the approximate pulse windows for the other antennas for each ray type can be calculated. The exact pulse arrival time in the other antennas is however not known: first of all, the electric field model used in the reconstruction does not contain phase information, leading to a small timing shift in the antennas at larger baselines due to relative shifts in the viewing angle. Additional timing shifts are caused by the error in the vertex reconstruction or uncertainties in the antenna positions and ice model.

Therefore, in order to identify the exact arrival time within each pulse window, another correlation between the reconstructed waveform and the data is performed. For low signal amplitudes, this results in a position determined by thermal noise fluctuations, which results in an overestimation of the power in these pulses. To avoid this, only pulses which exceed a signal-to-noise (SNR) ratio of at least 3.5, defined as half the maximum peak-to-peak voltage (within the pulse window) divided by the root-mean-squared noise voltage $\sigma_\mathrm{noise}$, are included in the fit. This is illustrated in Figure \ref{fig:pulse_selection}, right. Note that the pulses in the four deep Vpols of the phased array are always included, as well as the pulses of the two Hpols (H7,H8) directly above them, as these are close enough to allow the pulse arrival times to be determined directly from the phased array. Finally, antennas where the two expected ray types have overlapping pulse windows, causing the pulses to interfere, are excluded from the fit.

The $\chi^2$ statistic to be minimized is obtained by forward-folding the model prediction of noiseless signal waveforms with the appropriate in-ice propagation and detector response effects, and comparing this with signal data (or simulations), which include a contribution of thermal noise. If the noise is assumed to be approximately Gaussian, the $\chi^2$-statistic is given by
\begin{equation}
    \chi^2 = \sum_{n=1}^{n_{\textrm{pulses}}} \sum_{i=1}^{n_{\textrm{samples}}} \frac{(x_{i} - f_{i}(\theta_\mathrm{view}, \phi_\mathrm{pol}, E_{\mathrm{sh}}))^2}{\sigma_\mathrm{noise}^2},
\end{equation}
with $x_{i}$ the voltage data at sample $i$, $f_{i}$ the model prediction at sample $i$, $\theta_\mathrm{view}$ the viewing angle, $\phi_\mathrm{pol}$ the polarization angle and $E_\mathrm{sh}$ the shower energy. Only the $n_{\textrm{pulses}}$ that pass the SNR $>$ 3.5 criteria (with a minimum of $n_{\textrm{pulses}}$ = 6), with a window of  $n_{\textrm{samples}}$ are included. 

The pulses in the Hpol are typically shorter than the pulses in the Vpol, and therefore a smaller fitting time-window $\Delta t$ is used (30 ns) than for the Vpol (60 ns). This results in a different number of $n_{\textrm{samples}}$ (sampling rate $\cdot \Delta t$) depending on the antenna type. $\sigma_\mathrm{noise}$ is the noise root-mean-squared of the antennas after filters, which are applied to reduce the noise contribution in regions where the antenna is not sensitive. A low-pass filter of 700 MHz and a high-pass filter of 50 MHz are used, resulting in a typical $\sigma_\mathrm{noise}$ = 11.6 mV for this particular set-up.

\begin{figure}
    \centering
    \includegraphics[width=.49\textwidth]{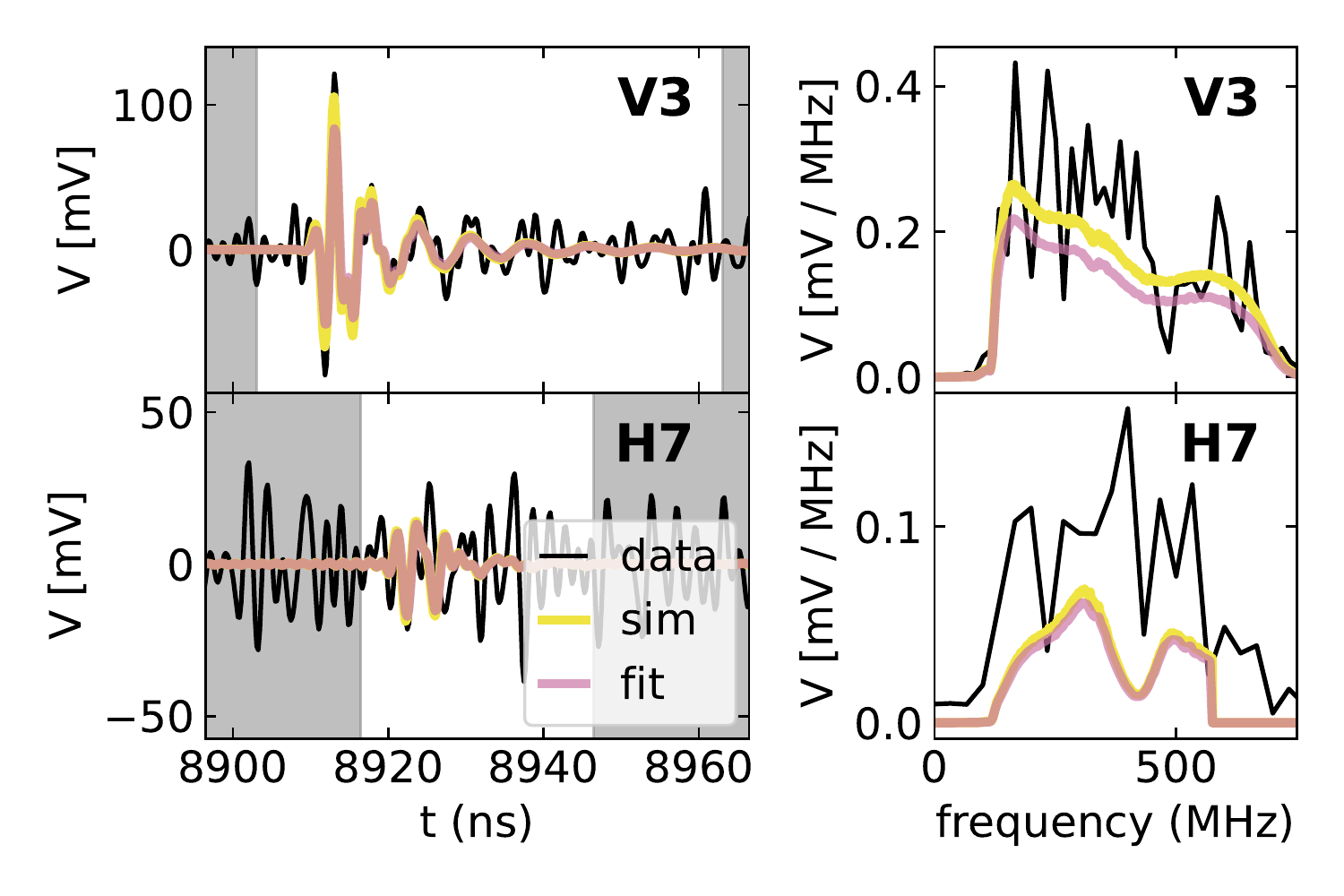}
    \caption{The reconstructed pulses for one Vpol (top) and one Hpol (bottom) antenna in an example event, compared to the noiseless (sim) signal. The non-shaded region in the waveform (left) indicates which part of the waveform is included in the fit. The shape of the frequency spectrum is dominated by the system response.}
    \label{fig:example_reco}
\end{figure}

The results of the fitting procedure for one example event and two of the included pulses are shown in Figure \ref{fig:example_reco} and compared to the true (simulated) neutrino signal. Note that the result here is additionally constrained by the fit to the 10 other pulses (not shown).

\section{Performance}
\label{sec:performance}
In order to test the reconstruction algorithm, we generate a representative set of neutrino signals as they would be detected in a deep radio station. We reconstruct the entire set and discuss the reconstruction performance for individual events, several high quality subsets, and the implications for radio detectors. 
\subsection{Simulation set-up}
\label{sec:simulationset}

We use NuRadioMC to generate $\sim$ $6.5 \cdot 10^7$ neutrinos with energies ranging from $10^{16} - 10^{22}$ eV, following a spectrum that combines an IceCube-like flux \cite{IceCube_flux_2017} with an additional contribution from cosmogenic neutrinos as defined in \cite{GZK}. The neutrino arrival directions are generated randomly, although their attenuation due to propagation in the Earth is taken into account by weighting the resulting events appropriately.

The simulated detector is based on the design of the RNO-G station, as described in \cite{RNO-G:2020rmc}. The power string and the helper strings (see section \ref{subsection:detector_response}) are separated horizontally by \SI{35}{m}. The power string consists of a phased array containing four Vpol antennas at $\sim$ \SI{100}{m} depth, with two Hpol antennas directly above, as well as 3 additional Vpol antennas at depths of 80, 60 and \SI{40}{m}. The two helper strings feature two Vpol antennas with one additional Hpol antenna directly above, at a depth of \SI{100}{m}. The antenna responses for the Hpol and Vpol are used as available in NuRadioMC. The biconal Vpol is sensitive between \SIrange[]{100}{600}{MHz} with a resonant frequency of about \SI{150}{MHz}. The Hpol is significantly lower in gain and has its highest response between \SIrange[]{200}{500}{MHz}. Both have a symmetric response in azimuth and are most sensitive at a zenith of 90$^\circ$. Both are connected to an amplifier (Low-Noise Amplifier, \cite[Figure~17]{RNO-G:2020rmc}) that boosts the signal strength (almost) frequency independent up to \SI{800}{MHz}.

We model the phased array trigger (four Vpols, \SI{1}{m} spacing) with a $2 \sigma$ threshold trigger for a single antenna, which is studied to be the equivalent for the trigger used by RNO-G as shown in \cite[Figure~18]{RNO-G:2020rmc}. A sampling rate of \SI{2.4}{GHz} is used, similar to the RNO-G design sampling rate. We simulate infinite trace lengths (trace lengths are adjusted such that all arriving pulses are stored). The influence of shorter recorded waveforms is not studied in this work. While no significant changes are expected for, e.g. the RNO-G hardware (record length of 850 ns), the impact could be larger for deeper stations as planned for IceCube-Gen2 Radio \cite{IceCube-Gen2:2020qha}. 

The noise due to thermal fluctuations of the electrons in the antennas can be described with
\begin{equation}
 \sigma_\mathrm{noise} =    \sqrt{k_b\Delta f T_{\textrm{noise}} 50 \Omega},
\end{equation}
with $k_b$ the Boltzmann constant, $\Delta f$ the frequency band, and $T_{\textrm{noise}}$ the noise temperature. We simulate the noise with $T_{\textrm{noise}}$= 300 K, corresponding to a $\sigma_\mathrm{noise} = 11.6\ \mathrm{mV}$ root-mean-squared of the noise within our frequency band of 50-700 MHz.  We simulate the noise in the frequency domain with amplitudes that are Rayleigh distributed, resulting in (almost) Gaussian distributed noise in the time-domain.

The electric field model of \cite{Alvarez-Muniz:2011wcg} for the Askaryan emission is used, as described in section \ref{subsection:signal_model}. The ice refractive index is modeled with a depth-dependent refractive index model, described with an exponential function, as suggested by \cite{PhysRevD.98.043010} and a depth and frequency dependent attenuation model from \cite{Avva:2014ena}. 

Note that the simulations performed are a statistical realization in two ways, i.e.\ 1) the thermal noise fluctuations and 2) the electric field generated from statistically varying shower profiles.

With this neutrino flux and detector configuration, we obtain an integrated weight of 4273 triggering events, of which 1881 contain only a hadronic shower. The reconstruction algorithm is run for each of these, and the results are described in the following section.

\subsection{Reconstruction performance}
\label{sec:rec_performance}
\begin{figure*}
	\centering
	\includegraphics[width=.95\textwidth]{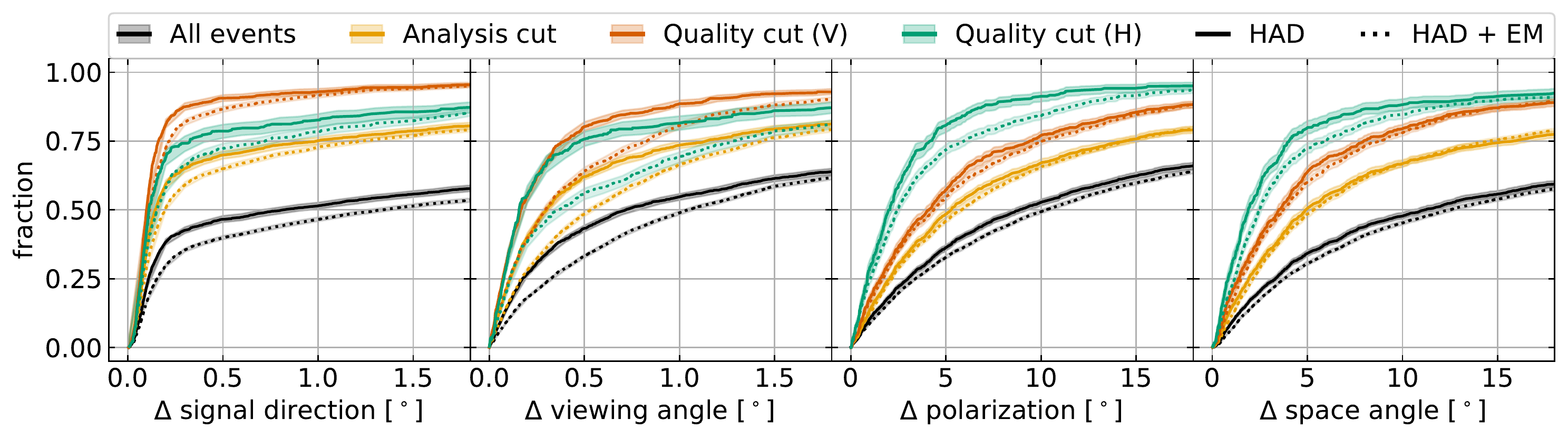}
	\caption{The cumulative distribution of the obtained resolution for different aspects of the reconstruction and different subsets of events. Note the different scales on the x-axis. Shaded regions indicate the approximate $1 \sigma$ statistical uncertainty.}
	\label{fig:individual_reconstructions}
\end{figure*}

The results of the reconstruction are shown in Figure \ref{fig:individual_reconstructions}. We distinguish the \textit{signal direction} resolution, i.e.\ the angle between the true and reconstructed emitted signals, the \textit{viewing angle}, \textit{polarization angle}, and \textit{space angle}, i.e.\ the angle between true and reconstructed neutrino directions. Of these, the first depends only on the performance of the reconstruction of the shower maximum. We show results both for hadronic events (solid lines), for which our code was originally developed, as well as those for all events (including electromagnetic interactions, dashed lines). Performance for this latter category is expected to be worse, due to the more irregular shape of electromagnetic showers experiencing the LPM effect, as well as potential interference between the emission from the electromagnetic and hadronic showers. Particularly for these events, it is expected that the performance could be improved by developing a dedicated algorithm and/or classifier, as has for example been investigated in a machine-learning approach in \cite{Glaser:2022lky}.

In addition to the results for the full event set, we show three different subsets of events that emphasize the different features of the reconstruction. These are selected by imposing a minimum cut on the SNR in certain antennas, here defined as the maximum amplitude of the true (noiseless) signal divided by the true noise root-mean-squared amplitude $\sigma_\mathrm{noise}$. The three subsets are:
\begin{itemize}
	\item The \textit{analysis cut}: this includes events that have a signal with an SNR of at least 2.5 in one phased array antenna, as well as at least one other antenna in the power string. In addition, a signal with an SNR of at least 2 is required in each of the helper strings. This cut primarily improves the quality of the shower-maximum reconstruction, which requires a signal in at least these four antennas (groups) to uniquely determine the origin of the emission. 
	\item The \textit{quality cut (V)}: this includes events that pass the analysis cut, and in addition have a signal with an SNR of at least 3.5 in one of the two uppermost Vpol antennas. This cut aims to further improve the quality of the shower-maximum reconstruction, by ensuring the presence of a clear signal at at least 5 different baselines. 
	\item The \textit{quality cut (H)}: this includes events that pass the analysis cut, and in addition have an SNR of at least 3 in at least one of the Hpol antennas. This leads to a significantly stronger constraint on the polarization angle, which is the dominant uncertainty in the space angle.
\end{itemize}
The percentage of events retained by each of these cuts is shown in Table \ref{tab:cutefficiencies}.

\begin{table}
    \centering
    \begin{tabular}{c|rr|rrr}
        &\multicolumn{2}{c|}{$n_\mathrm{events}$}&\multicolumn{3}{c}{Cut efficiency [\%]}\\
        $\log_{10}\frac{E_\nu}{\mathrm{eV}}$&had&all&A.C.&Q.C.\ (V)& Q.C.\ (H)\\
        \hline
        16.0-16.5 & 130& 496&   6&$<$1&$<$1\\
        16.5-17.0 & 242& 755&  33&   5&   2\\
        17.0-17.5 & 383& 945&  54&  23&  12\\
        17.5-18.0 & 458& 897&  64&  37&  14\\
        18.0-18.5 & 319& 608&  73&  47&  17\\
        18.5-19.0 & 223& 360&  79&  54&  28\\
        19.0-22.0 & 125& 212&  90&  62&  34\\
        \hline
        \textbf{16.0-22.0}&\textbf{1881}&\textbf{4273}&\textbf{  59}&\textbf{  33}&\textbf{  15}
    \end{tabular}
    \caption{The (weighted) number of triggered events, with the percentage of events retained by the analysis cut (A.C.) and the quality cuts (Q.C.) introduced in section \ref{sec:rec_performance}, binned in neutrino energy $E_\nu$. The efficiencies shown are for hadronic events only, though the (binned) efficiencies are similar for all events.}
    \label{tab:cutefficiencies}
\end{table}

The obtained resolution is limited by the shower-maximum reconstruction. When this step is successful, the signal direction is reconstructed with a median resolution of $0.1-0.2 ^\circ$. However, in particular for low-SNR signals, or shallow vertices, which experience more significant ray bending in the firn, there are long tails in the distribution corresponding to misreconstructed vertices. This leads to wrong signal windows (i.e.\ the fit may not even include the actual pulses) and an incorrect geometry, which in general cause the rest of the reconstruction to fail, too.

If the shower maximum is reconstructed successfully, the dominant uncertainty is the polarization angle. The viewing angle is a priori constrained to lie within a couple of degrees of the Cherenkov angle, and can be reconstructed relatively well depending on the amplitude of the signal in the Vpol antennas. On the other hand, partly because of the relatively weaker response of the Hpol antennas, as described in section \ref{subsection:detector_response}, the resolution of the polarization angle is about an order of magnitude worse than that of the viewing angle. This leads to a space angle resolution that closely resembles the polarization angle resolution. As expected, the biggest improvement here is observed by applying a cut on the minimum SNR in the Hpol antennas (Quality cut (H)).

The difference between purely hadronic events and those including electromagnetic showers is most clearly visible in the signal direction and viewing angle resolutions. This is mostly due to two factors; firstly, the larger spread in the shapes of the individual showers for electromagnetic showers, and secondly, the current reconstruction algorithm uses an analytic model for the Askaryan spectrum of hadronic showers, which is slightly different from that for electromagnetic showers. A dedicated algorithm that additionally targets electromagnetic showers will likely improve on these results.

\begin{figure}
	\centering
	\includegraphics[width=.49\textwidth]{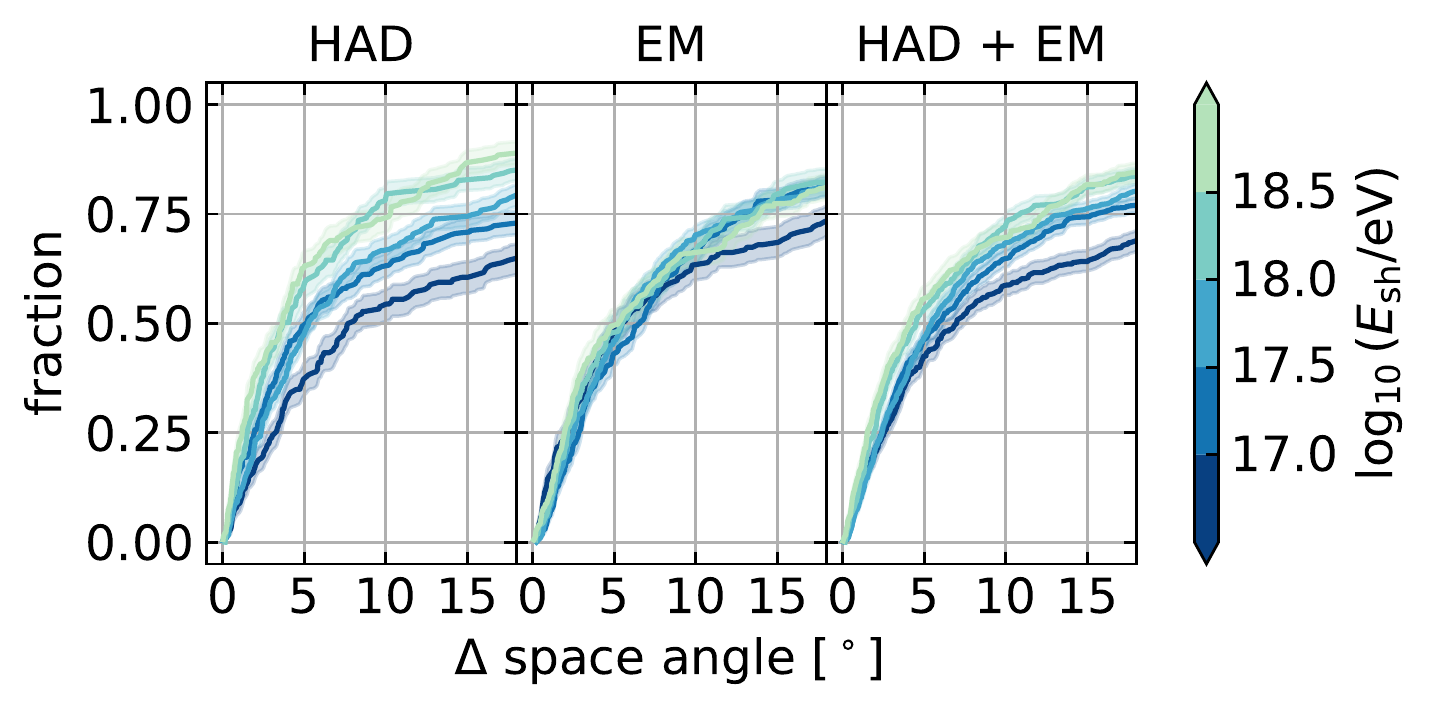}
	\caption{The space angle distribution for events passing the analysis cut, for different shower types and energies. For hadronic events, high-energy showers in general give higher amplitude signals, leading to an improving resolution as a function of energy.}
	\label{fig:space_angle_vs_energy}
\end{figure}

\begin{figure}
	\centering
	\includegraphics[width=.49\textwidth]{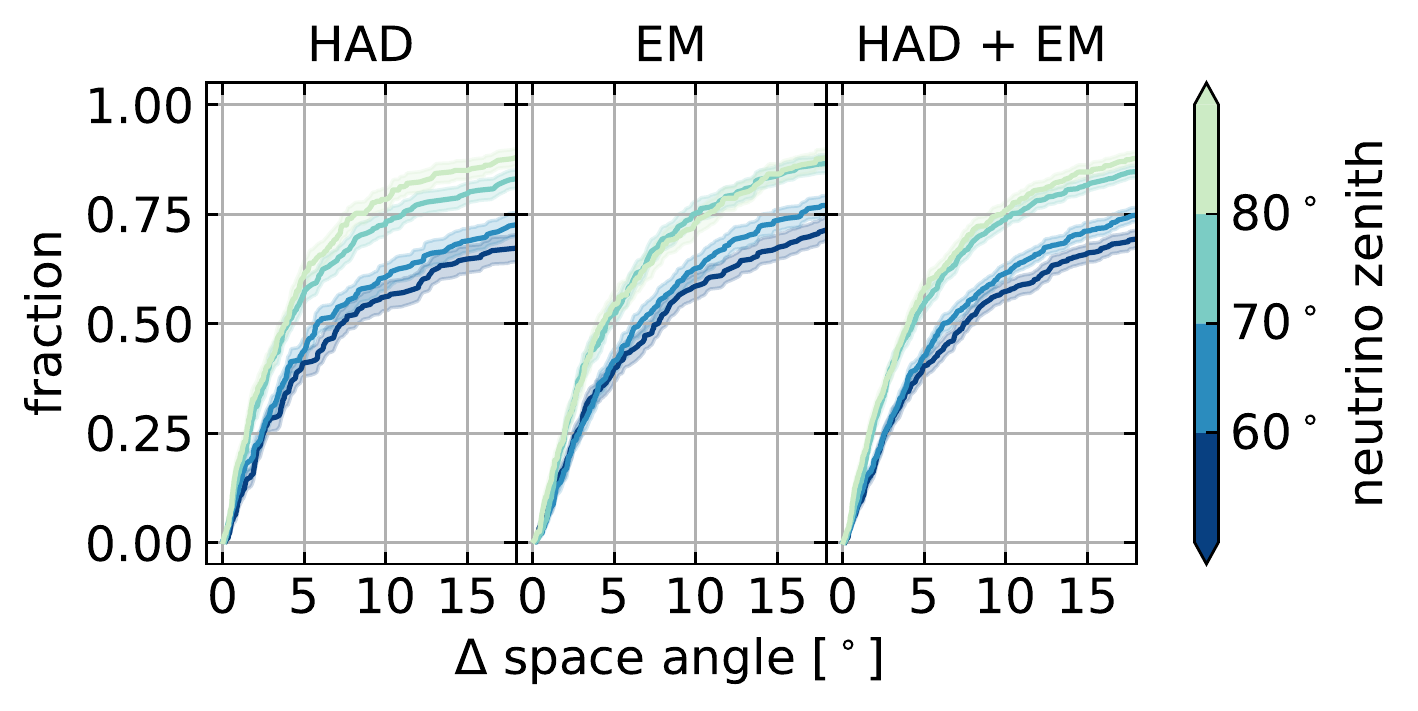}
	\caption{The space angle distribution for events passing the analysis cut, for different shower types and zenith bins. More vertical showers correspond to shallower interaction vertices, due to the emission angle of approximately $56^\circ$, which in turn are harder to reconstruct.}
		\label{fig:space_angle_vs_zenith}
\end{figure}

The resolution additionally varies with the deposited shower energy and the neutrino zenith. The energy dependence for the space angle for events passing the analysis cut is shown in Figure \ref{fig:space_angle_vs_energy}. In general, even after the cuts are applied, higher energy showers are usually higher in amplitude in both Hpol and Vpol antennas, leading to an improving resolution as a function of shower energy. However, for electromagnetic showers, this effect is almost completely offset by the increased spread in the shower shape due to the LPM effect at higher energies. Furthermore, for lower energies a slightly better resolution in space angle for electromagnetic showers is obtained than for hadronic showers. 

The dependence on the neutrino zenith angle is shown in Figure \ref{fig:space_angle_vs_zenith}, again for events passing the analysis cut only. In this case, the resolution improves as the neutrino becomes more inclined (larger zenith). There are two reasons for this. Firstly, more vertical showers ($< 60^\circ$) are geometrically constrained, by the Cherenkov angle of $\sim 56 ^\circ$ and the fact that rays bend downwards in the firn, to lie closer to the surface, which leads to a more challenging shower-maximum reconstruction, as mentioned previously. In addition, as the polarization vector points towards the shower axis, more inclined neutrinos include a stronger horizontally polarized component, which improves the reconstruction of the polarization angle. These effects are similar for both hadronic and electromagnetic events.

\begin{figure}
	\centering
	\includegraphics[width=.49\textwidth]{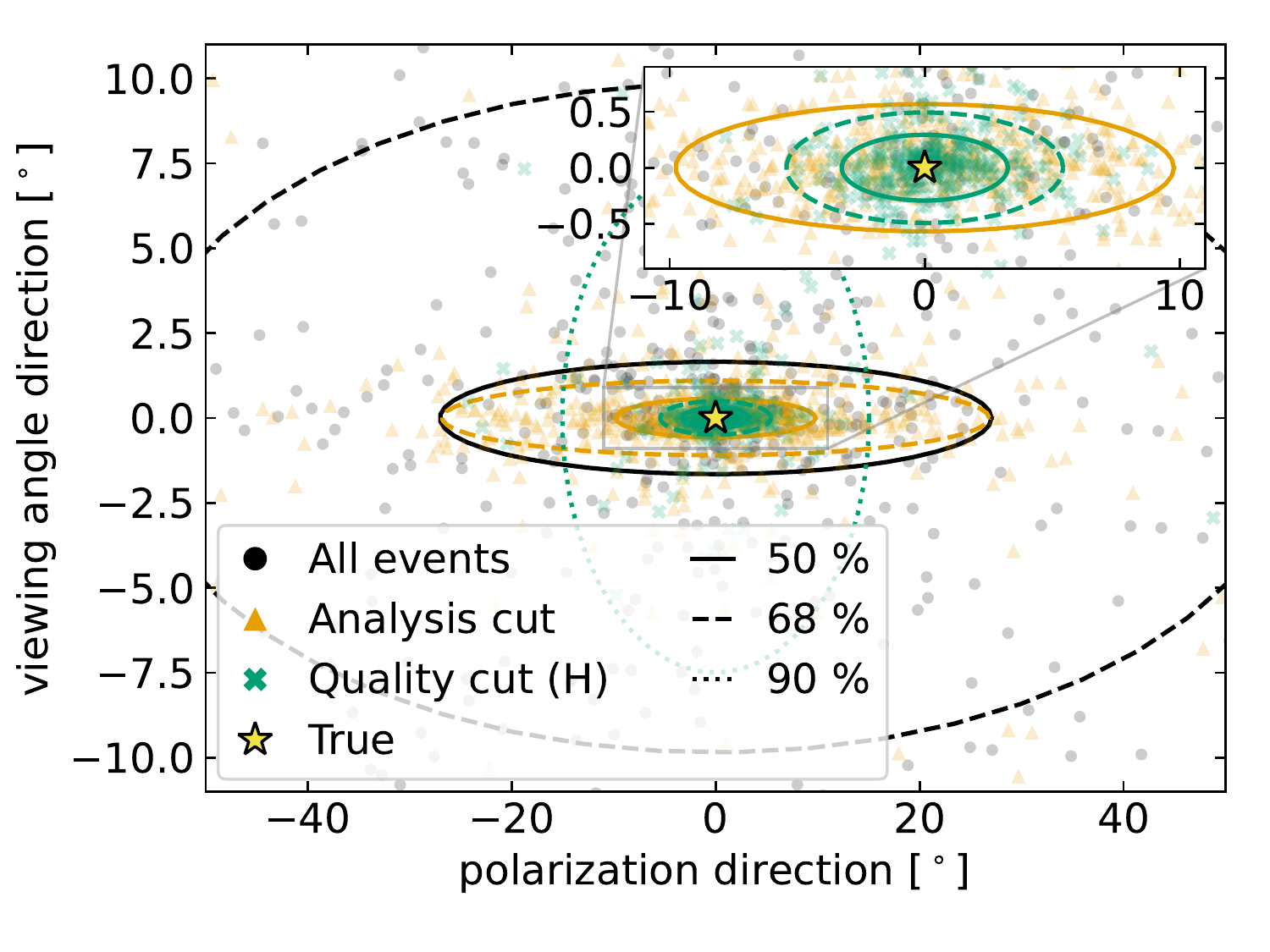}
	\caption{Distributions of the reconstructed neutrino directions in two dimensions, rotated and projected into a coordinate frame where the polarization angle points along the x-axis and the viewing angle along the y-axis. Shown are individual events, as well as contour lines for different cuts, for all hadronic events.}
	\label{fig:2d_all_events}
\end{figure}

Finally, it is important to keep in mind that the resolution, in terms of space angle, is almost completely dominated by the polarization angle, and therefore can not be straightforwardly interpreted as the relevant characteristic for e.g.\ point source searches. It is however possible to consider the reconstructed events in a rotated and projected reference frame, such that the (large) uncertainty in the direction of the polarization angle lies along the horizontal axis, and the (much smaller) uncertainty along the viewing angle direction along the vertical axis (c.f.\ Figure \ref{fig:educative_figure}, top left). The resulting asymmetric distribution is shown in Figure \ref{fig:2d_all_events}. For the smaller quantiles, the contours are strongly elongated along the polarization direction, with the size of the contour decreasing with analysis and quality cuts. The larger quantiles, however, include events with poor shower-maximum reconstruction, for which the reconstruction error is more random, leading to more symmetric contours (e.g.\ the 90\% contour for quality cut (H)). 

\begin{figure}
    \centering
    \includegraphics[width=.49\textwidth]{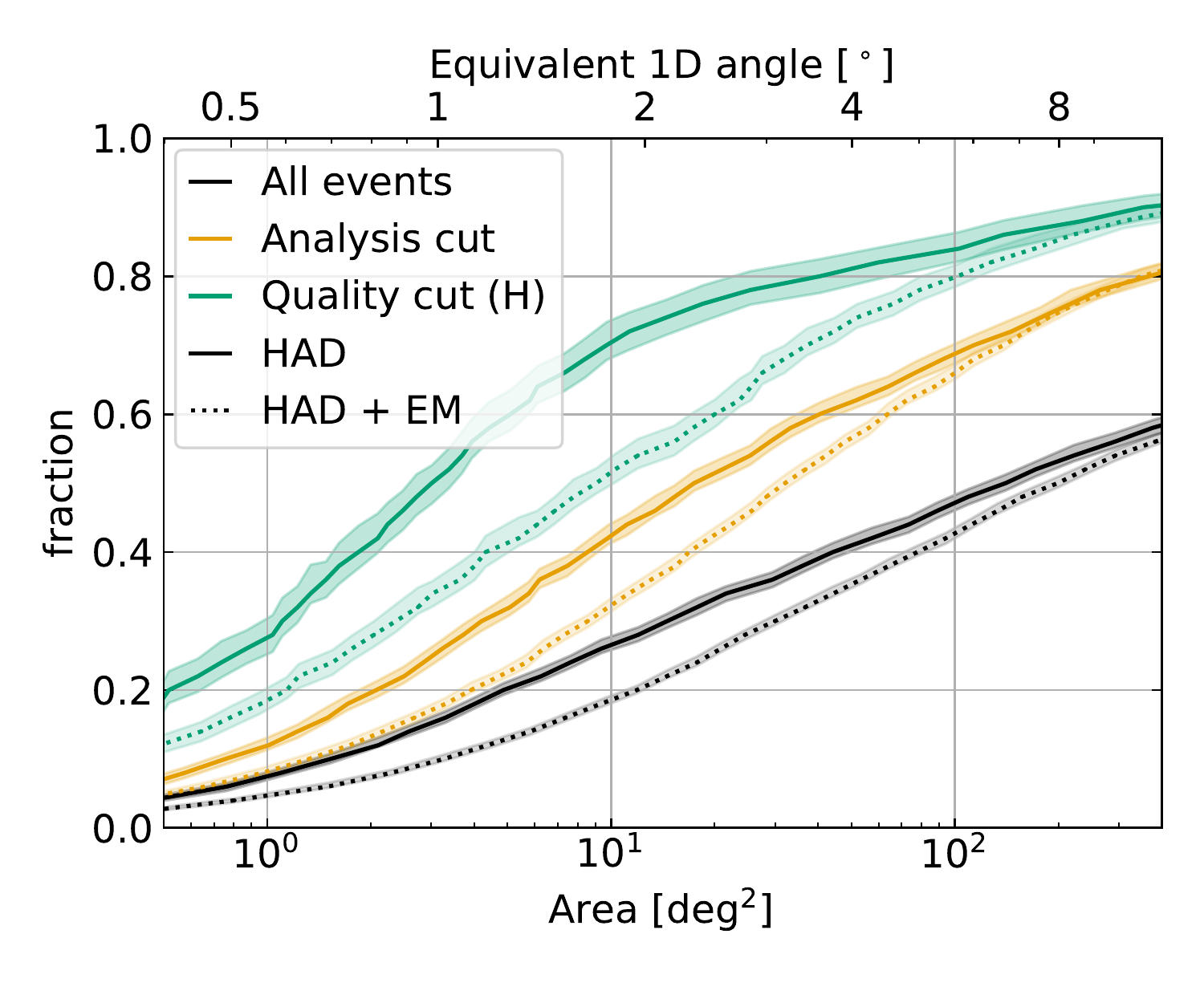}
    \caption{The cumulative distribution of the obtained resolution, in terms of area in square degrees. The secondary axis indicates the angle that would correspond to the same surface angle in the case of a symmetric angular distribution. Shaded regions indicate the approximate $1 \sigma$ statistical uncertainty. 
    }
    \label{fig:square_degrees}
\end{figure}

We can then compute a resolution in square degrees, corresponding to the surface of an ellipse covering N\% of events. This is done in Figure \ref{fig:square_degrees}. The secondary x-axis shows the corresponding 1D angle that would cover the same area for a symmetric contour (i.e.\ a circle). The median 2D resolution for hadronic events passing the analysis cut, for example, is approximately $17\ \mathrm{deg}^2$, corresponding to a 1D angle of $2.4 ^\circ$, compared to a median space angle of $4.9^\circ$. This figure also emphasizes the importance of improving the viewing angle reconstruction for electromagnetic events; although the resolution in terms of space angle is very similar for most events, including electromagnetic events increases the median spread in 2D by a factor of 2-3 in area.

\subsection{Event contour and point spread function}
\label{sec:contours_data}

\begin{figure}
    \centering
    \includegraphics[width=0.49\textwidth]{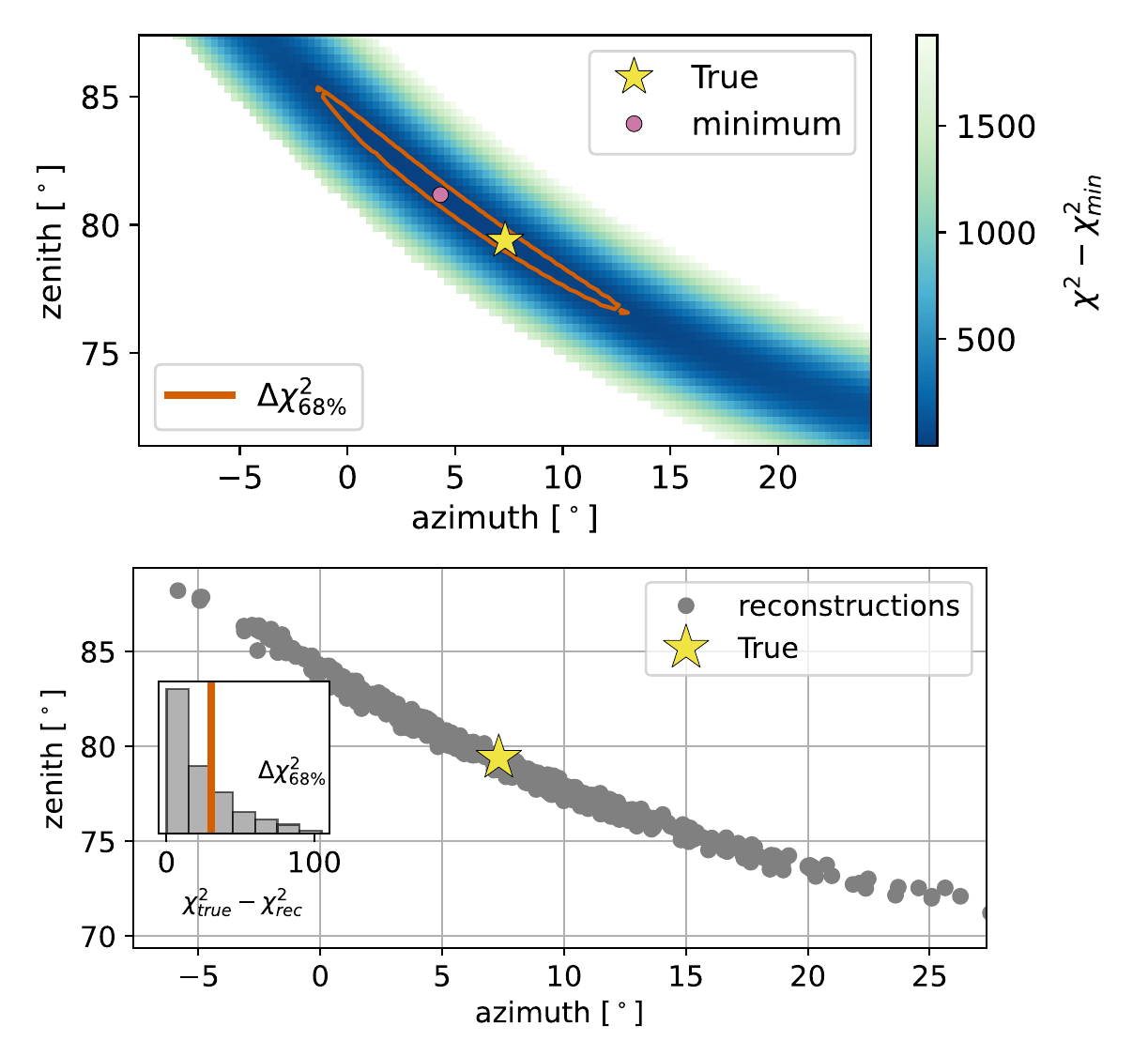}
    \caption{Top: The 68$\%$-contour for an event of average quality that passes the analysis cuts, resulting in a space angle of 4$^\circ$. The uncertainty region is an elongated and narrow asymmetric area. Bottom: Distribution of the reconstructed direction for the same event under the influence of statistical uncertainties, e.g.\ shower profile and thermal noise fluctuations.}
    \label{fig:contours}
\end{figure}

Next, we study uncertainty contours for individual events. Figure \ref{fig:contours} shows the $68\%$ uncertainty region for an event that passed the analysis cuts. 
The contour is obtained as follows: on a grid in zenith and azimuth around the best fit point the test statistic is computed and a 68\% contour of constant $\Delta\chi^2 = \chi^2 - \chi^2_\textrm{min}$ is drawn. 
The appropriate value of $\Delta \chi^2$ is calculated by reconstructing the same event multiple times with different realizations of the electric field and the thermal noise, and determining the 68$^\textrm{th}$ percentile of $\chi^2_{\textrm{true}} - \chi^2_{\textrm{rec}}$, where $\chi^2_{\textrm{true}}$ is the test statistic corresponding to the true direction and best fitting shower energy. Thus, the contour can be interpreted such that 68\% of the repeated experiments will be reconstructed within this contour. This is shown in the bottom plot of Figure \ref{fig:contours}.

\begin{figure}
    \centering
    \includegraphics[width=0.45\textwidth]{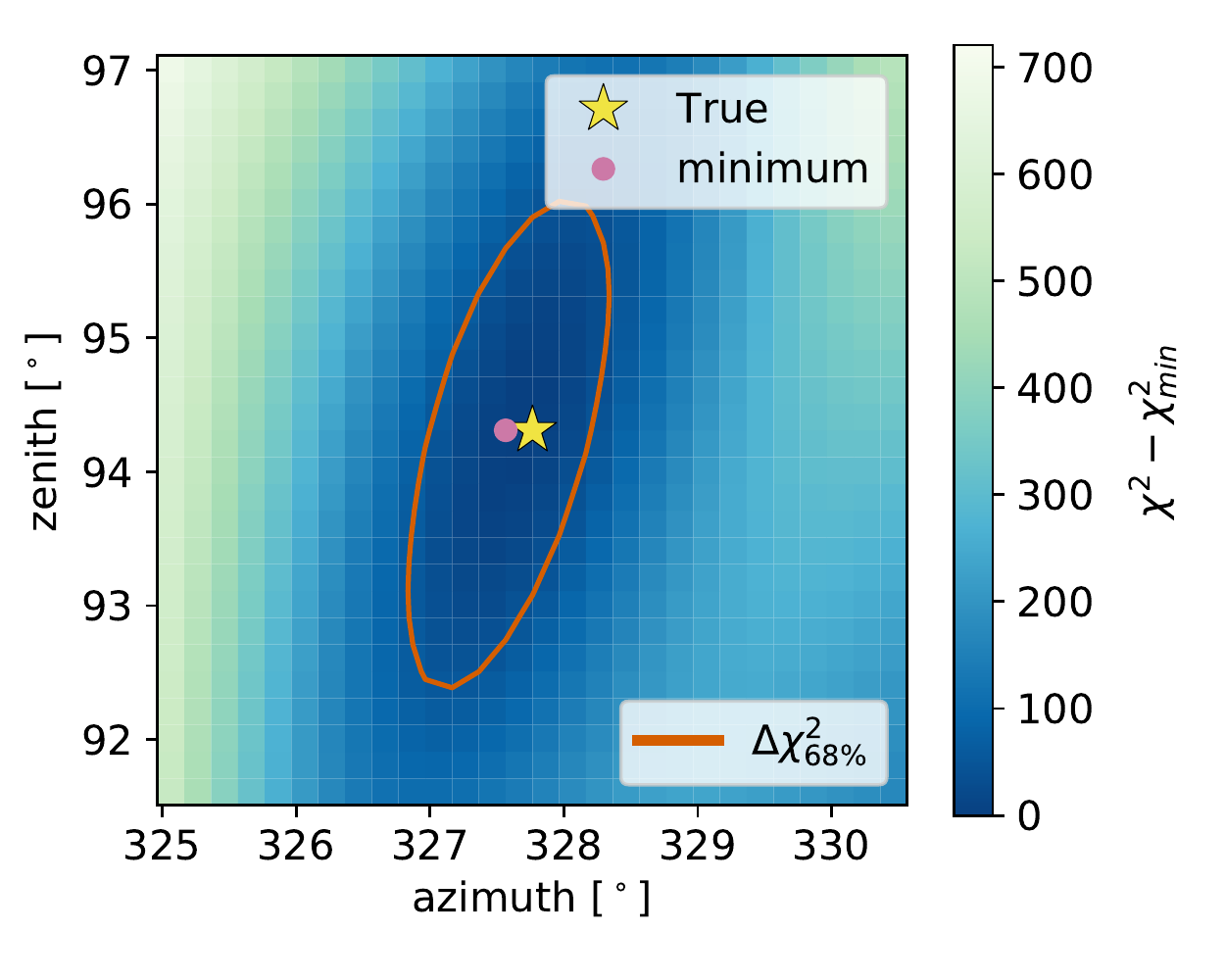}
    \caption{Same as Figure \ref{fig:contours} (top panel), but for a high-quality event with large contribution in the Hpol antennas.}
    \label{fig:contour_Hpol}
\end{figure}

An example contour for an event with a large amplitude in the Hpol is shown in Figure \ref{fig:contour_Hpol}. A $68\%$ area of 4.5 deg$^2$ is obtained. As can be seen from the figure, for small polarization angles the uncertainty region in zenith and azimuth can be closely approximated by an ellipse. 

The event contours for neutrinos detected with the deep station, even those of high quality, are highly asymmetric, and therefore we emphasize again that quoting the difference in angle between best fit point and true direction as the angular resolution is somewhat misleading. 

\begin{figure}
    \centering
    \includegraphics[width=0.45\textwidth]{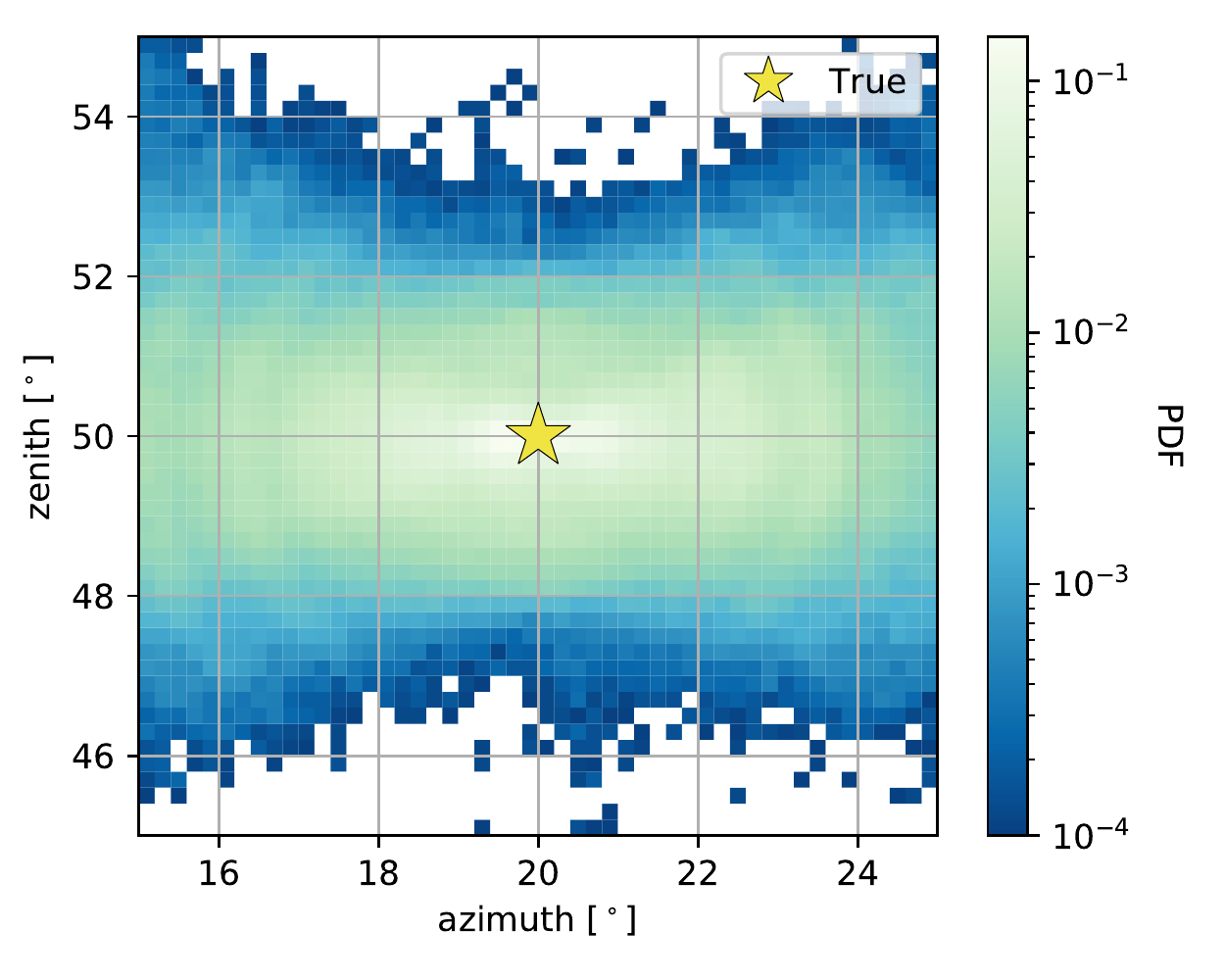}
    \includegraphics[width=0.45\textwidth]{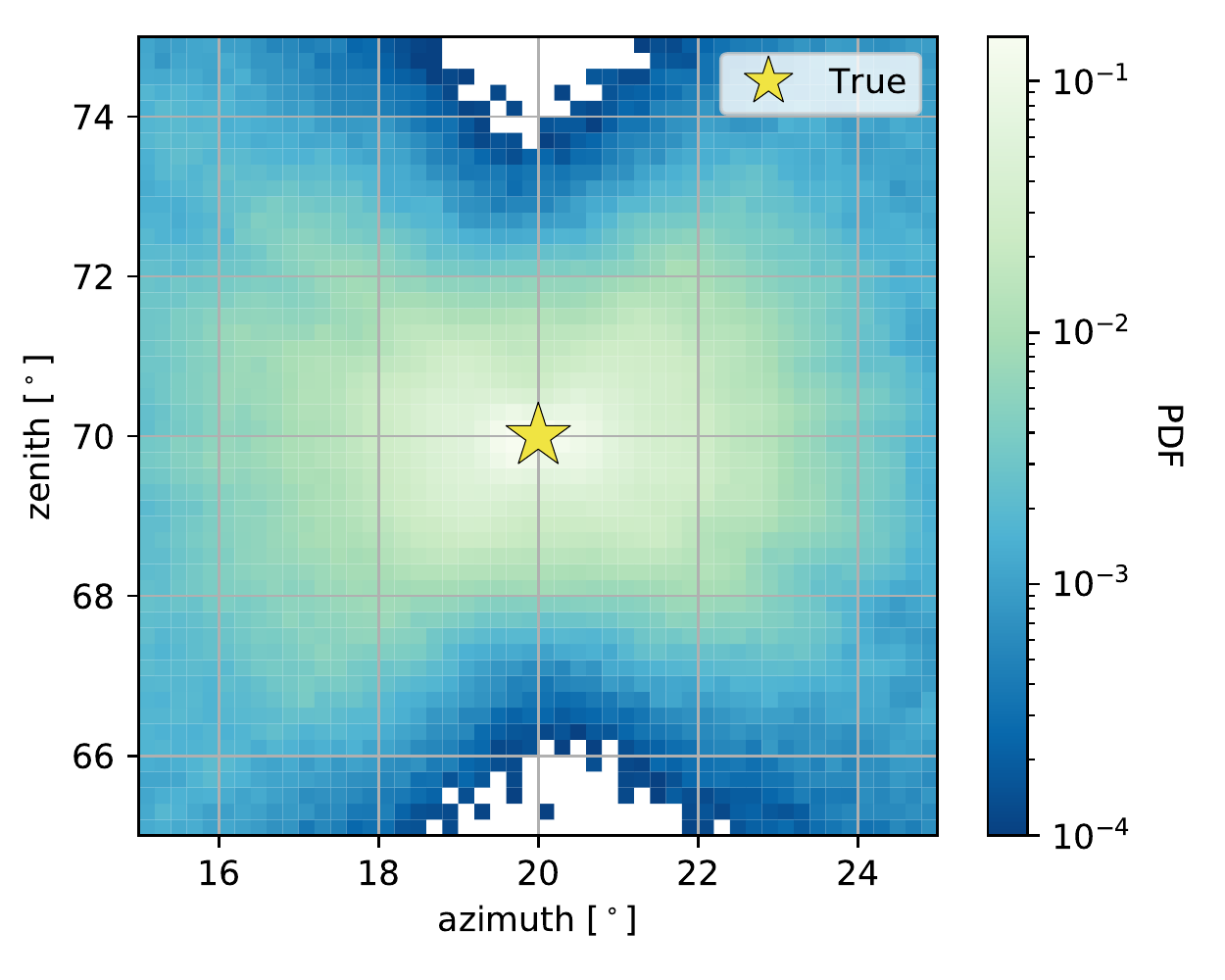}
    \caption{Top: simulated best-fit points using the obtained resolution for quality cut (H) of the uncertainties in signal direction, viewing angle and polarization for a point source located at zenith of 50$^\circ$. For details refer to the text. Bottom: same as the top figure, but for a source located at zenith of 70$^\circ$. Clearly seen is the zenith dependent shape of the point spread function.}
    \label{fig:psf_sim}
\end{figure}

The orientation of the polarization direction depends on the signal direction, which is different from the neutrino direction, resulting in a 'bow-tie'-shaped PSF for a neutrino point source, as explained in section \ref{sec:contours}. The geometry of the radio cone, as well as the Vpol-based trigger, limit the allowed orientations of the polarization.

In order to estimate the shape of the PSF, ideally one would want to know the full 3-dimensional probability distribution in signal direction, viewing angle and polarization. Alternatively, one could simulate the distribution of reconstructed events for a given fixed source position. As either option would require a dedicated effort of significant computational cost, we instead proceed as follows. First, we divide the distribution into two subsets based on the quality of the signal direction: 'well reconstructed' ($\Delta < 2^\circ$) and 'poorly reconstructed' ($\Delta > 2^\circ$). The resolutions for either subset are then assumed to be independent between signal direction, viewing angle and polarization, which we know to be a good approximation for individual events. The PSF is then obtained by folding the resulting resolutions with the distribution of allowed true signal directions for a specific local zenith. We expect the resulting PSF shape to qualitatively agree with the shape that would have been obtained with a dedicated study, and sufficiently accurate to highlight the features of interest.

The PSF for two hypothetical sources, at $50^{\circ}$ and $70^{\circ}$ local zenith, are shown in Figure \ref{fig:psf_sim}. Only events (hadronic and electromagnetic) passing quality cut (H) are included. For larger incoming zenith angles of the neutrino, a broader part of the cone becomes observable and thereby a larger range of polarization angles. Consequently, the PSF becomes broader and the forbidden region shrinks. Note that these are 'instantaneous' PSFs. Except at the poles, a source at fixed declination and right ascension will in general appear at different local zeniths, leading to a slightly larger and more 'smeared' time-integrated PSF. However, as long as the detection times are available, using the instantaneous PSF is more accurate and powerful in a source search.

The area of the PSF containing N\% of events can be estimated by 'pixel counting'. As the above procedure results in a space angle distribution that is slightly different from the true distribution, we correct for this by multiplying the area by the ratio $(\Delta\psi_\mathrm{true}/\Delta\psi_\mathrm{PSF})^2$, where $\Delta\psi$ denotes the space angle, at each quantile. 
The resulting areas lie somewhere in between the area expected for a single event and the area of a symmetric PSF, and additionally depend on the local zenith, as demonstrated. For a source at a zenith of 70$^\circ$ (Figure \ref{fig:psf_sim} bottom, all events (hadronic + electromagnetic) passing quality cut (H)) the median space angle is 2.5$^\circ$, which would imply a median area of 19$^\circ$ for a symmetric PSF. The actual PSF area obtained is 14 deg$^2$ at 70$^\circ$. At a zenith of 50$^\circ$, there are two competing effects - the PSF is flatter, but the reconstruction quality is lower (see Figure \ref{fig:space_angle_vs_zenith} and the discussion there). This results, overall, in a larger area than at 70$^\circ$, albeit with large uncertainties due to the low statistics of events passing quality cut (H) at 50$^\circ$. 

\begin{figure}
    \centering
    \includegraphics[width=0.49\textwidth]{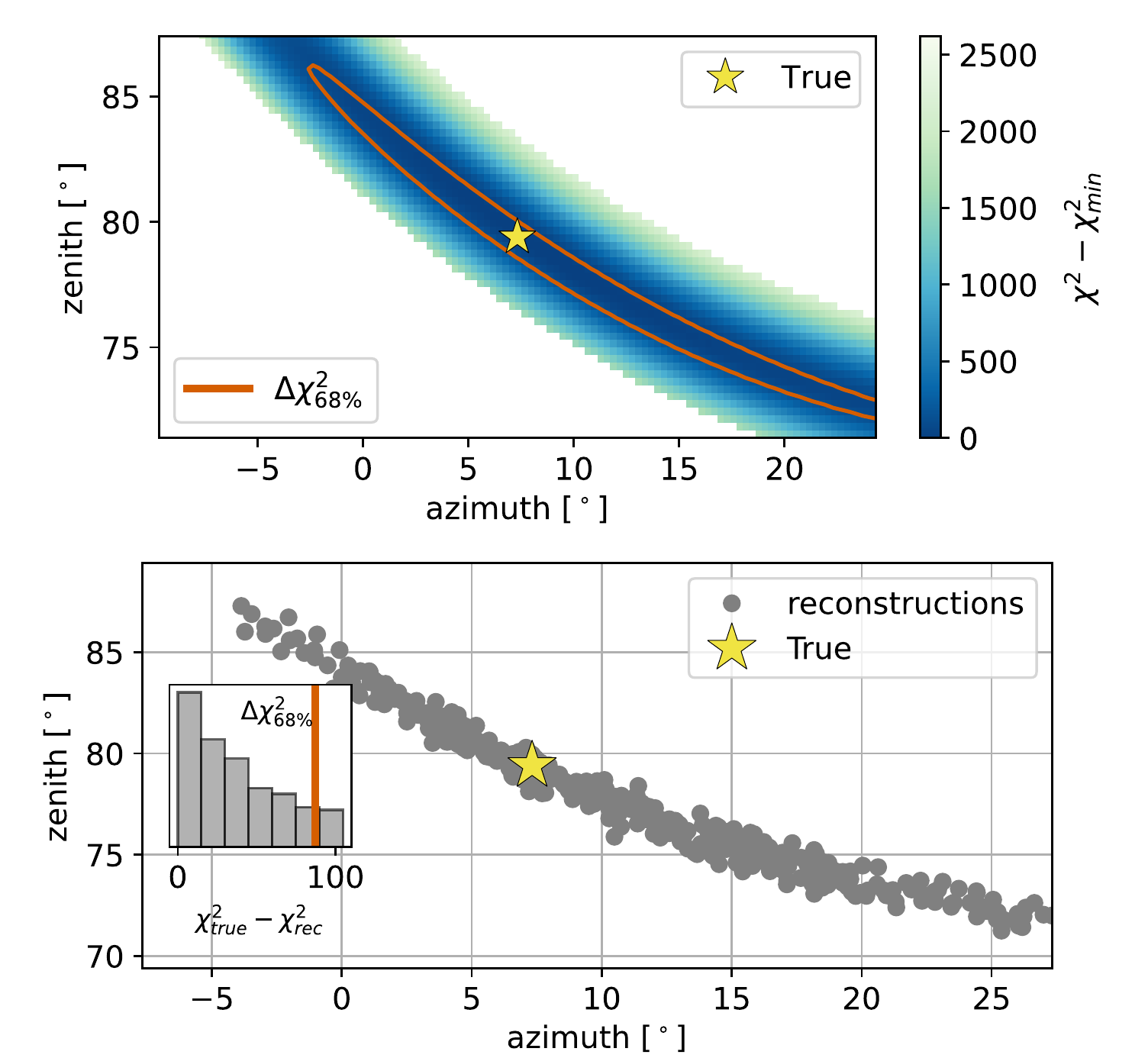}
    \caption{Same event as Figure \ref{fig:contours}, including systematic uncertainties. }
    \label{fig:contour_sys}
\end{figure}

\subsection{Systematic uncertainties}
\label{systematic}
To demonstrate the effects of an imperfect knowledge of the detector and the ice target volume for the outlined reconstruction method, we compute an event contour for the same example event shown in Figure \ref{fig:contours}, i.e.\ a typical event remaining after analysis cuts. We take into account three systematic uncertainties, i.e.\ 1) antenna positions, 2) antenna response and 3) the ice refractive index model. We include an offset in the position of the antenna of \SI{3}{cm} in any direction, and uncertainty of 10\% in the antenna vector effective length and a \SI{10}{MHz} uncertainty in the frequency behavior of the vector effective length (i.e.\ a shift in the resonance frequency and the phase behavior). We use Gaussian uncertainties for each individual antenna. For the ice model, we let the parameters $\Delta n$ and $z_{0}$ of the exponential refractive index profile $n(z) = n - \Delta n\cdot e^{\frac{z}{z_{0}}}$ vary by $10\%$, as the deep refractive index can be established very accurate.
While further work is needed to correctly treat the systematic uncertainties, we believe that these assumptions provide a reasonable estimate of the current understanding of radio neutrino detectors. 

The main impact of the systematic uncertainties for this example event is a worsening of the polarization resolution, from $\sigma_{68\%}$ = 9.2$^\circ$ to $\sigma_{68\%}$ = 11.5$^\circ$. Additionally, we obtain a bias of 4$^\circ$ in the reconstructed polarization. This is due to the fact that at low SNR, the arrival time at the Hpol antenna is determined from the arrival times in the nearby Vpol antennas. Uncertainties in the ice model or antenna positions result in a small timing offset between the reconstructed and true pulses, and therefore a systematic underestimation of the Hpol contribution. 
We emphasize the challenge of the low Hpol signals and state the importance of the development of algorithms to better identify low signal amplitudes in the thermal noise in future work. 

The resulting event contour is shown in Figure \ref{fig:contour_sys} resulting in an increase from 13 to 35 deg$^2$ for the example event. The impact of the mismatch of the model is observed in broader $\chi^2_{\textrm{min}} - \Delta \chi^2$ distribution (inset Figure \ref{fig:contour_sys} bottom).   

\subsection{Implications for instrument capabilities}

A good pointing resolution for the detected signals is important for neutrino identification, the diffuse flux discovery, cross-section measurements, and source identification. In this section we focus on the latter to illustrate the effect of analysis cuts and angular resolution.

Neutrino sources can be identified in two ways; either through a (temporal and spatial) coincidence with the detection of a photon in the electromagnetic spectrum or a gravitational wave, e.g.~\cite{IceCube:2018cha,Stein:2020xhk}, or by the detection of an excess of neutrinos that is large enough to identify it above the expected background, e.g.~\cite{NGC}. There are varying analysis strategies for temporal coincidences, where for instance in a very short coincidence window, also a limited angular resolution can lead to a significant detection. However, it is in general understood that a smaller angular uncertainty will always lead to a better chance of finding true coincidences in multi-messenger searches and a reliable arrival direction reconstruction is a prerequisite to a successful triggering of follow-up observations at other wavelengths. 

For illustration, we focus here on a potential point source excess, since it is most strongly affected by the angular resolution. A good angular resolution reduces the expected number of background events within the uncertainty region (PSF) of the neutrino (source) and therefore reduces the flux required for source identification. 

Identifying point sources by detecting an excess of events above the expected background and implications for the large scale in-ice radio array of IceCube-Gen2 has been extensively studied in \cite{Fiorillo:2022ijt}. We note that for smaller radio experiments such as RNO-G we rely on very extreme flux models from point sources or nearby sources to be able to identify the source. Nonetheless, we study the effect of the different analysis cuts introduced in section \ref{sec:performance}, to establish the best search strategy.

We use the software framework toise \cite{vanSanten:2022wss}, which allows for the fast calculation of upper limits and discovery potentials of neutrino telescopes based on parameterizations of the detector response in terms of effective volume, angular resolution and energy resolution. The fast performance is due to the usage of an \textit{Asimov} dataset, meaning that the observed event rates are replaced by the exact mean \cite{Cowan:2010js}. Additionally, the test statistic of the null hypothesis (background only) is assumed to be $\chi^{2}$-distributed, i.e.\ Wilk's theorem holds \cite{Wilks:1938dza}. The behavior of the test statistic for the  background only scenario can be computed by creating multiple pseudo-experiments using the observed background rates. Background events for UHE neutrinos have not been observed and are as uncertain as the UHE neutrino flux itself, and therefore this approach is as realistic as possible at this stage.

As background for identifying point sources of UHE neutrinos we include three kind of events; 1) the tail of the high-energy neutrino flux, 2) cosmogenic neutrinos, and 3) UHE muons. 
We include for 1) the diffuse neutrino flux as observed by IceCube, assuming a power-law extension in neutrino energies without a cut-off \cite{IceCube_flux}, 2) a cosmogenic neutrino flux resulting from cosmic rays interacting with photon fields assuming a 10$\%$ proton fraction for the cosmic-ray spectrum observed by the Pierre Auger Observatory \cite{GZK}, and 3) in-ice penetrating UHE muons stemming from cosmic-ray air showers inducing electromagnetic showers while radiating \cite{muon}. Each of these contributions are rare and likely on the same level of the expected number of UHE neutrinos, but uncertainties are large. For 1) the flux is measured at low energies (PeV range) with a set of spectral uncertainties, which results in large variations for the EeV regime even if the power-law assumption without cut-off holds. The uncertainties for 2) are dominated by the uncertain UHE cosmic-ray mass composition which influences the neutrino rate prediction by orders of magnitude.  Finally for 3) the number of predicted UHE muons in air showers varies due to large uncertainties on the hadronic interaction models, in particular at the highest energies. Therefore, we use the above mentioned background rate as our benchmark background and show results for varying background levels.

Other potential backgrounds for UHE radio detectors are currently under study and are not included in our background estimates. These backgrounds include radio emission from air showers penetrating into the ice and not fully developed in-ice penetrating air showers inducing electromagnetic dense shower cores. Both backgrounds, like the muons, can be reduced if the corresponding air shower can be detected, for example with a shallow component of the hybrid station. We, however, do not include such a cosmic-ray veto in the background assumption. 

The impact of the different event groups introduced in section \ref{sec:performance} for different background levels is studied in terms of the $5 \sigma$ discovery potential, i.e. the source flux required for source identification, for a steady point source (10 years) at 20$^\circ$ declination assuming a neutrino flux of $E^{-2}$. 

Results are shown based on the angular resolution obtained in this work, an energy resolution based on \cite{Aguilar:2021uzt}, the effective volume based on the specifics of RNO-G in Greenland, using a station effective volume for an RNO-G station and an array of 35 stations \cite{RNO-G:2020rmc}. 20$^\circ$ declination is directly in the field of view of a radio detector based in Greenland. 

The point source study modeled with toise is searching 20$^\circ$ around a source. The framework assumes 1D symmetry for the resolution of a source, i.e.\ the number of background events is calculated in a circle with radius of the angular resolution. Therefore, the angular resolution required is the PSF, i.e.\ how a point source is perceived given the imperfect detector. We convert the PSF, obtained as explained in section \ref{sec:contours_data}, into the area-equivalent symmetric resolution for a source located at a local zenith of 70$^\circ$. For a station based in Greenland a source at a declination of 20$^\circ$ has an average local zenith of 70$^\circ$, which is therefore a good approximation for the size of the PSF.

\begin{figure}
    \centering
    \includegraphics[width=0.49\textwidth]{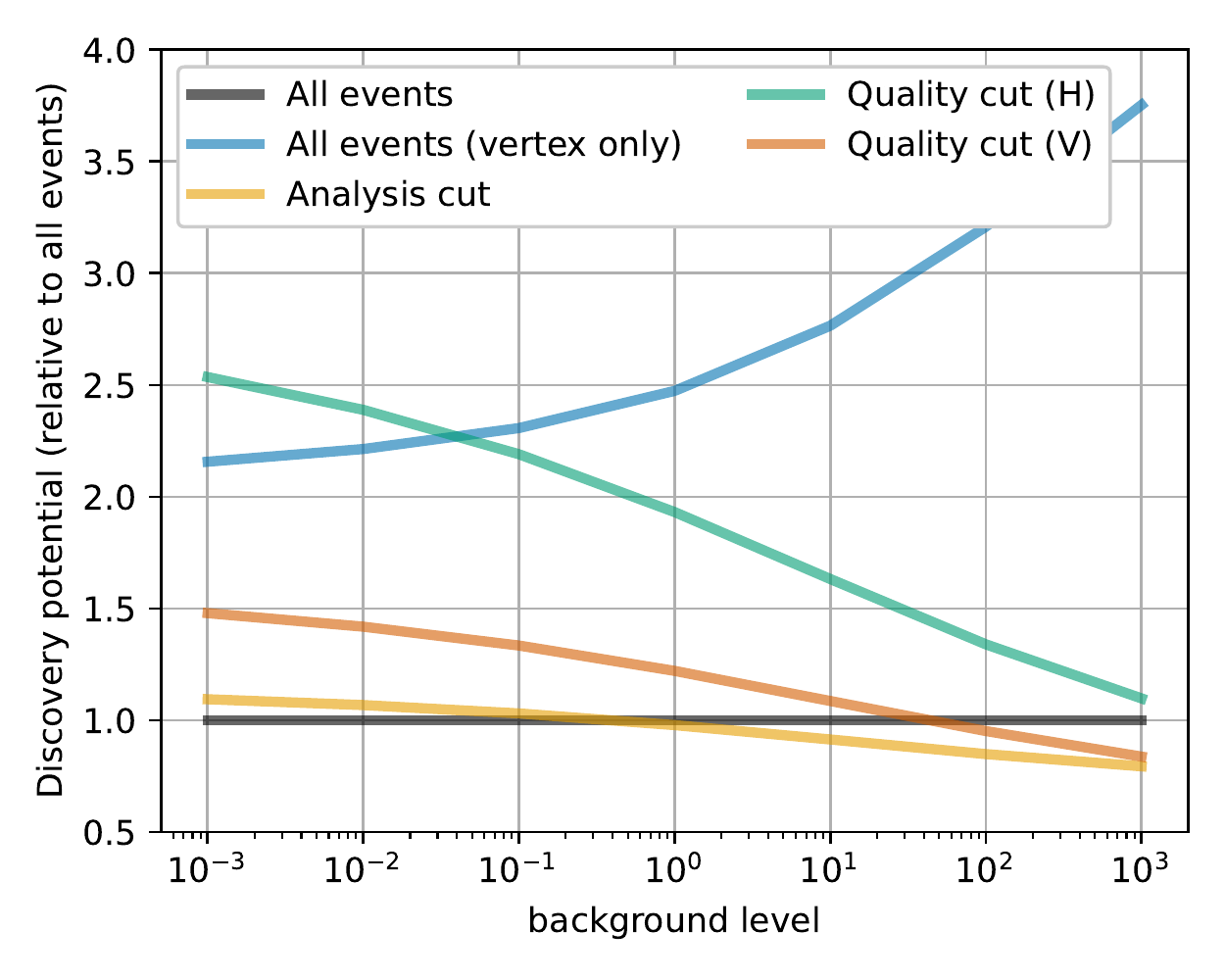}
    \caption{Fractional improvement of the $5 \sigma$ discovery potential for different analysis cuts compared to all events with the obtained angular information for a declination of 20$^\circ$. Note that an event group has a better performance than the benchmark set when the relative discovery potential is lower. We show the analysis cut and two quality cuts (V and H), as discussed in section \ref{sec:rec_performance}. Furthermore, we show the relative discovery potential for the full set using only the shower-maximum resolution. Details are given in the text.}
    \label{fig:toise_result}
\end{figure}

Our results are visualized in Figure \ref{fig:toise_result}. We show the ratio of the discovery potential for different event groups, improving in angular resolution while reducing the effective volume with their corresponding angular resolution compared to the benchmark event set (full set with obtained resolution). In addition to the event groups introduced in section \ref{sec:performance}, we also show the results for the full event set with only the shower maximum reconstructed, i.e.\ the resolution obtained when only exploiting the signal direction. We include this to show the improvement provided through this work. 

We conclude that the obtained neutrino direction resolution results in a factor 2-3 improvement in discovery potential compared to reconstructing the shower maximum only. At very low background fluxes, the discovery potential depends only mildly on the number of background events, and thus depends more strongly on the effective volume than the resolution of each event set. As the background flux increases, the reduction in background allowed by restricting to the analysis cut, or eventually one of the quality cuts, becomes favorable, compared to including all events in the analysis. Note, again, that the discovery potential is shown \emph{relative} to that of all events; for lower background levels or for transient searches the overall discovery potential improves.

\section{Conclusion}
\label{sec:conclusions}

In this article, we have described a full reconstruction algorithm for deep in-ice radio neutrino detectors based on the forward-folding principle. We have shown how the signal direction, viewing angle, and polarization combine to uniquely define the neutrino direction. The performance of the algorithm was quantified for each of these parameters, and it was demonstrated that the much larger resolution for polarization leads to strongly asymmetric uncertainty contours for single events. Furthermore, as the alignment of different event contours depends on the interaction vertex of the neutrino rather than its direction, the PSF for a neutrino source in turn exhibits a (zenith-dependent) 'bow-tie' shape, rather than resembling that of a single event contour. We therefore emphasize that the 1D space angle distribution is not sufficient to adequately describe the reconstruction resolution.

The discovery potential for a point source (for a 20$^\circ$ search) with this algorithm is shown to improve by a factor $\approx$ 2 compared to reconstructing the vertex position only. The resolution obtained is shown to depend on the signal strength in the different antennas. For hadronic events, a large subset of events ($\sim 60\%$) can be selected in order to obtain a median resolution of $4.9^\circ$ (space angle) or $17\ \mathrm{deg}^2$ ($\approx2.4^\circ$ 1D equivalent). With stricter cuts, a smaller median resolution is obtained at the cost of effective volume. In terms of discovery potential, however, we demonstrate that the loss of statistics generally outweighs the potential improvement in resolution, even for relatively optimistic flux assumptions.

The reconstruction algorithm described in this article is included as part of the open-source NuRadioMC framework, enabling end-to-end simulation and performance studies and optimization for deep in-ice radio neutrino detectors. In future work, improvements are to be expected by improving the identification of pulses with a small signal-to-noise ratio, for which the current strategy of cross-correlation occasionally leads to a failure to correctly identify the pulse position or even to a misreconstruction of the vertex position. In addition, although the performance of the reconstruction for electromagnetic showers is already similar in some aspects (e.g.\ polarization), further gains are to be expected from a dedicated algorithm for this type of events.

\begin{acknowledgements}
We are thankful to our colleagues from the RNO-G and the IceCube-Gen2 collaboration for both controversial and constructive discussions about angular reconstruction for radio detectors.  

We acknowledge funding from the German research foundation (DFG) under award NE 2031/2-1, which has enabled this study.
\end{acknowledgements}

\bibliographystyle{JHEP}
\bibliography{BIB}

\end{document}